\documentclass[lettersize,journal]{IEEEtran}
\usepackage{amsmath,amsfonts}
\usepackage{algorithmic}
\usepackage{algorithm}
\usepackage{array}
\usepackage[caption=false,font=normalsize,labelfont=sf,textfont=rm]{subfig}
\usepackage{textcomp}
\usepackage{stfloats}
\usepackage{url}
\usepackage{verbatim}
\usepackage{graphicx}
\usepackage{bm}
\usepackage{cite}
\usepackage[font=footnotesize]{caption}
\usepackage[font=footnotesize]{subcaption}  % 控制子图标号(a)(b)字体

\begin{document}

\title{Semantic Communication with an LLM-enabled Knowledge Base}

\author{Wuxia Hu, Caili Guo, \emph{Senior Member, IEEE}, Yang Yang, \emph{Senior Member, IEEE}, \\
Chunyan Feng, \emph{Senior Member, IEEE},
Kuiyuan Ding and Shiwen Mao, \emph{Fellow, IEEE}
\thanks{
An earlier version of this paper
was presented at IEEE International Conference on Communications (ICC) 2025.  

W. Hu, C. Guo, C. Feng and K. Ding are with the Beijing Key Laboratory of Network System Architecture and Convergence, School of Information and Communication Engineering, Beijing University
of Posts and Telecommunications, Beijing 100876, China (e-mail: wuxiahu@bupt.edu.cn; guocaili@bupt.edu.cn; cyfeng@bupt.edu.cn; dingkuiyuan@bupt.edu.cn).

Y. Yang is with the Beijing Laboratory of Advanced Information Networks, School of Information and Communication Engineering, Beijing University of Posts and Telecommunications, Beijing 100876, China (e-mail: yangyang01@bupt.edu.cn).

S. Mao is with the Department of Electrical and Computer Engineering, Auburn University, Auburn, AL 36849, USA (email: smao@ieee.org).
}
}

\markboth{Journal of \LaTeX\ Class Files,~Vol.~14, No.~8, August~2021}%
{Shell \MakeLowercase{\textit{\textit{\textit{et al}}.}}: A Sample Article Using IEEEtran.cls for IEEE Journals}


\maketitle

\begin{abstract}

Semantic communication (SC) can achieve superior coding and transmission performance based on the knowledge contained in the semantic knowledge base (KB). 
However, conventional KBs consist of source KBs and channel KBs, which are often costly to obtain data and limited in data scale. 
Fortunately, large language models (LLMs) have recently emerged with extensive knowledge and generative capabilities.
Therefore, this paper proposes an SC system with LLM-enabled knowledge base (SC-LMKB), which utilizes the generation ability of LLMs to significantly enrich the KB of SC systems. In particular, we first design an LLM-enabled generation mechanism with a prompt engineering strategy for source data generation (SDG) and a cross-attention alignment method for channel data generation (CDG).
However, hallucinations from LLMs may cause semantic noise, thus degrading SC performance. 
To mitigate the hallucination issue, a cross-domain fusion codec (CDFC) framework with a hallucination filtering phase and a cross-domain fusion phase is then proposed for SDG. In particular, the first phase filters out new data generated by the LMKB irrelevant to the original data based on semantic similarity. 
Then, a cross-domain fusion phase is proposed, which fuses source data with LLM-generated data based on their semantic importance, thereby enhancing task performance.  Besides, a joint training objective that combines cross-entropy loss and reconstruction loss is proposed to reduce the impact of hallucination on CDG.
Experiment results on three cross-modality retrieval tasks demonstrate that the proposed SC-LMKB can achieve up to 72.6\% and 90.7\% performance gains compared to conventional SC systems and LLM-enabled SC systems, respectively.
\end{abstract}


\section{Introduction}

\IEEEPARstart{W}{ith} the rapid development of sixth-generation (6G) networks, the Internet of Things (IoT) stands out as one of the most important application domains\cite{guo2021enabling}. IoT refers to the network of interconnected devices that collect and exchange data to enable smart applications across various sectors, including security, transportation, and urban management. However, conventional communication methods, which prioritize bit-level precision, are increasingly insufficient to meet the growing demands of IoT systems. The transmission of large volumes of data from numerous devices leads to low communication efficiency and high bandwidth requirements, posing significant challenges in managing the massive scale of data and devices in IoT environments \cite{yang2022semantic}. 

To address the challenge, semantic communication (SC) emerges as a potential key technology in the 6G\cite{yang2022semantic, liu2024ofdm, jascyang}. SC aims to extract semantic features from the source and then perform various tasks at the receiver rather than merely ensuring the accurate transmission of symbols.
By extracting the essential semantics from data, SC effectively reduces the transmission of redundant information. A typical SC system mainly consists of the semantic codec and the semantic knowledge base (KB). In recent years, SC has been increasingly explored in conjunction with multiple-input multiple-output (MIMO) systems, leveraging their spatial diversity and multiplexing gains to improve semantic transmission performance\cite{zhang2024scan,xie2024robust,xie2023communication}. To effectively extract the semantics from the source and interpret the received semantics at the receiver, the SC system significantly relies on contextual knowledge provided by the semantic KB\cite{yang2022semantic, wang2022performance}. In the context of SC, semantic KB is defined as a well-structured model with powerful processing, memory, and reasoning capabilities to provide rich knowledge in supporting semantic coding and transmission\cite{ren2024knowledge}. Specifically, the KB enables the semantic encoder to obtain semantic features based on the task objectives and allows the semantic decoder to interpret the received representation with respect to shared knowledge in the semantic KB rather than relying solely on the raw signal.

Existing KBs in SC systems can be generally categorized into source KBs and channel KBs \cite{ren2024knowledge}. 
Specifically, source KBs store semantic knowledge related to the source information, such as labeled source datasets, knowledge graph (KG) and feature codebooks, which are used to enable semantic codec and transmission. Most existing works focus on source KBs \cite{xie2021deep,yi2023deep,zhang2022deep,luo2022semantic,zhou2022cognitive,liukg,shi2021new}. At its early stage, most semantic source KBs are primarily composed of labeled source datasets \cite{xie2021deep,yi2023deep,zhang2022deep,luo2022semantic}. To improve the interpretability of the semantic KBs, some works constructed KBs by knowledge graph (KG) and feature codebooks extracted from the source data \cite{zhou2022cognitive,liukg,shi2021new}. Although the aforementioned traditional KB systems have demonstrated satisfactory performance, obtaining high-quality annotated data is often challenging due to the high cost of data labelling, and thus the size of datasets can be limited, especially for complex tasks in IoT scenarios\cite{SKB, li2024data}.
For example, in cross-modal retrieval tasks within smart IoT environments, SC systems rely on well-constructed source KBs to enable the semantic codec to accomplish the retrieval task. However, building such KBs requires large-scale paired data across modalities, which is expensive and labor-intensive to obtain. The scarcity of high-quality labeled data significantly limits the construction of the source KBs and then the accuracy of retrieval performance in IoT SC systems.

On the other hand, channel KBs face the same data scarcity challenge \cite{ren2024knowledge,tang2025efficient}. In particular, channel KBs contain channel-related data, including CSI, signal-to-noise ratio (SNR), and other wireless channel characteristics, which support adaptive semantic transmission over varying channel conditions. 
For example, Tang \textit{et al}.\cite{tang2025efficient}. proposed a channel KB constructed by CSI index values to support efficient CSI feedback in MIMO systems, enabling the transceivers to share a common understanding of the channel representation. 
In the context of IoT SC systems, channel-related data plays key role in facilitating semantic transmission. For instance, accurate CSI data is essential for precoding and beamforming to combat channel fading in MIMO systems. However, obtaining CSI data is challenging since traditional pilot-based methods incur significant transmission overhead, which is often impractical in large-scale IoT networks.
Overall, existing KBs, including source KBs and channel KBs, lack the capability to solve the data-limited challenge in complex tasks, which significantly limits the performance of SC systems \cite{jiang2024large}. Therefore, semantic KBs need to be significantly enhanced, especially in data-limited scenarios.

 \subsection{Related Works and Challenges}
We first introduce the related works in SC with semantic KBs, which can be categorized into SC with source KBs and SC with channel KBs. Then, we discuss the challenges in the existing works. 

\textit{ 1) SC with source KBs}:
Within the area of source KBs, SC systems can be further divided into SC with conventional source KBs and SC with LLM-enabled source KBs.
In the early stages of SC, source KBs are primarily composed of labeled datasets \cite{xie2021deep,yi2023deep,zhang2022deep,luo2022semantic}. 
For example, Xie \textit{\textit{et al}}.\cite{xie2021deep} constructed a textual KB utilizing training datasets for text transmission, which can be shared with transceivers to train the semantic codec.
Then, to reduce transmission overhead, Yi \textit{\textit{et al}}.\cite{yi2023deep} constructed a textual KB by extracting partial sentences in text datasets and transmitting the residual information rather than the source information with the aid of the shared KB.
For image transmission, Zhang \textit{\textit{et al}}.\cite{zhang2022deep} considered an image KB with the empirical image data, which can convert the observed image data into a similar form of the empirical data in the original KB without retraining through transfer learning.
 
To enhance the interpretability of KBs, some works constructed KBs by KG\cite{zhou2022cognitive,liukg,shi2021new}. For text transmission, Zhou \textit{\textit{et al}}.\cite{zhou2022cognitive} uses text triples (including head entity, relationship, and tail entity) to describe semantic information. For text tasks, Liu \textit{\textit{et al}}.\cite{liukg} utilized KG as the semantic KB to extract task-related triplets from the source, which reduces transmission overhead and improves task performance. For speech transmission, Shi \textit{\textit{et al}}.\cite{shi2021new} extracted semantic features from the speech source by a KG-enabled KB. 
Recently, feature codebooks\cite{fu2023vector} and mapping relationships\cite{yang2022semantic} have also been regarded as a type of KB.
Fu \textit{\textit{et al}}.\cite{fu2023vector} utilized a discrete feature codebook as the semantic KB by quantifying source features into indices, and thus the proposed SC system can be compatible with digital communication systems. 
Yang \textit{\textit{et al}}.\cite{yang2022semantic} proposed a task-related KB for image classification, which is constructed by the relationship between image features and classification labels. However, due to the high cost of labeled data, the size of the dataset
can be limited \cite{jiang2024large}, especially for complex tasks. Therefore, existing KBs need to be significantly enhanced, especially in data-limited scenarios. 
  
 Due to the limited reasoning ability of conventional KBs, recent studies have explored the integration of LLMs as KBs. LLMs, such as Llama \cite{touvron2023llama} and DeepSeek \cite{liu2024deepseek}, have drawn widespread attention due to their powerful reasoning and generation ability \cite{LLM}. 
 Therefore, it can be used to enhance the semantic KBs in SC systems to acquire additional data and improve system performance. Recently, several studies have utilized the reasoning ability of LLMs as semantic KBs to extract semantic features\cite{guo2023semantic, jiang2024large}. For example, Guo \textit{et al}.\cite{guo2023semantic} employed LLMs as external KBs to evaluate semantic importance for text transmission. Meanwhile, Jiang \textit{et al}. \cite{JIANG} developed an LLM-enabled KB to extract personalized semantics from text. Besides, Yang \textit{et al}.\cite{generativeLLM} utilized the chain-of-thought (CoT) capability of LLMs to enable hierarchical semantic parsing for the targeted task and promote adaptive sequential semantic extraction to accommodate dynamic communication environments. 
 However, research on utilizing the data generation ability of LLMs as semantic KBs remains unexplored. 

 \textit{ 2) SC with channel KBs}:
Channel KBs contain channel-related data, including CSI, SNR, and other wireless channel characteristics \cite{ren2024knowledge}.
Channel KBs are typically constructed through pilot-based channel estimation procedures, where known reference signals are transmitted to probe the wireless environment. 
By providing transceivers with prior knowledge about the wireless channel, channel KBs can effectively assist in SC systems such as beamforming and optimization. 
For example, Tang \textit{et al}.\cite{tang2025efficient}. proposed a channel KB constructed by CSI index values to support efficient CSI feedback in MIMO systems, enabling the transceivers to share a common understanding of the channel representation. However, they still rely heavily on pilot-based estimation, which can introduce transmission costs and limit scalability in bandwidth-constrained environments. Thus, recent works have explored LLMs to generate channel-related data.
Liu \textit{\textit{\textit{et al}}}.\cite{liu2024llm4cp} proposed to finetune a pre-trained LLM to generate downlink CSI data based on the uplink data in both time-division duplex (TDD) and frequency-division duplex (FDD) systems. Then, Liu \textit{\textit{\textit{et al}}}.\cite{liu2025llm4wm} utilized a mixture of experts with low-rank adaptation to fine-tune LLMs to accomplish multi-channel tasks, including channel estimation and beam management. Besides, Fan \textit{\textit{\textit{et al}}}.\cite{fan2025csi} proposed a CSI prediction model based on LLMs to align the CSI modality with LLMs and then generate CSI data based on the history data. Although LLM-enabled methods have demonstrated the feasibility of adapting LLMs for channel KBs, they directly fine-tune LLMs to adapt to channel-related data, which deteriorates the natural language generation ability \cite{lobo2024impact}. Thus, existing works fall short of supporting data generation in both source KBs and channel KBs.

While LLMs hold great potential for generating diverse and complex data, existing works on LLMs neglect the impact of hallucinations from LLMs. LLMs often suffer from hallucinations  \cite{hallucination}. These hallucinations, where LLMs generate data that is inconsistent or entirely fabricated, can lead to semantic noise, thereby degrading the accuracy and relevance of the generated data. Consequently, the effective utilization of LLMs for data generation requires addressing the hallucination challenge to ensure the reliability and correctness of the generated content. Therefore, it is crucial to develop a more effective mechanism that can integrate LLMs into SC systems while resisting the semantic noise caused by hallucinations. To fully exploit LLMs as KBs in SC systems, two critical challenges must be addressed
 \begin{itemize}
     \item \textit{Challenge 1: How to effectively integrate LLMs into SC systems to accomplish data generation in both source KBs and channel KBs?}
     \item \textit{Challenge 2: How to resist the semantic noise introduced by hallucinations in LLMs?}
 \end{itemize}

\subsection{Contributions}
The main contribution of this paper is a SC system with an LLM-enabled KB (SC-LMKB) to address the aforementioned challenges. In particular, an LLM-enabled KB (LMKB) is employed to enable both SDG and CDG, where SDG focus on text data generation and is preliminarily introduced in our conference version \cite{huwuxia2025}. Besides, CDG focus on channel data generation by aligning CSI features with the natural language modality in the LLM space. Compared to our previous work \cite{huwuxia2025}, here, we propose a new CDG module to address the challenge of channel data acquisition.
To the best of the authors’ knowledge, this is the first work that introduces LLMs as semantic KBs to accomplish data generation in SC systems. The main contributions of this paper are summarized as follows
\begin{itemize}
\item We propose SC-LMKB that employs an LMKB to generate both data in both labeled source data and channel-related data based on the history data. In particular, a unified LLM backbone is used to alleviate data limitation issue in both source and channel KBs, enabling SC systems to enrich training data and reduce transmission overhead. The generated data can significantly enhance the semantic encoding and transmission performance of SC systems.
\item To address \textit{Challenge 1}, we propose an LLM-enabled generation mechanism, which leverages the LMKB to jointly generate data. In particular, the mechanism consists of two modules: SDG module and CDG module.
Specifically, for the SDG, we focus on text data generation where a prompt engineering strategy is proposed to enable LMKB to generate additional data. For the CDG, we propose a cross-attention alignment method that aligns CSI features with the natural language modality in the LLM space. This design allows us to leverage a unified LLM backbone for data generation in both source and channel KBs without requiring additional fine-tuning.

\item To address \textit{Challenge 2}, a cross-domain fusion codec (CDFC) framework is proposed to alleviate the hallucination in SDG. In particular, the framework contains a hallucination filtering phase to filter out the additional data irrelevant to the source based on semantic similarity. Then, a cross-domain fusion phase is proposed to utilize semantic importance to fuse source data with LLM-generated data, thereby enhancing task performance. Besides, we propose a joint training objective that combines cross-entropy loss and reconstruction loss to reduce the impact of hallucination on CDG. The joint training objective enables the model to simultaneously learn token-level semantics and signal-level structures for accurate CDG.

\end{itemize}

We apply the proposed SC-LMKB to a TPR task on three datasets to demonstrate its effectiveness. Experiment results show that the proposed SC-LMKB can achieve up to 72.6\% and 90.7\% performance gains compared to conventional SC systems and LLM-enabled SC systems, respectively.

The rest of this paper is organized as follows. Section II introduces the typical SC system model. Section III presents the proposed SC-LMKB. Section IV introduces the hallucination mitigation method. The experimental results and analysis are discussed in Section IV. Finally, Section V concludes this paper.

\section{System Model}
This section considers a typical SC system, where the semantic source data is transmitted over a wireless channel to facilitate downstream tasks. Then, we discuss the challenges of the typical SC systems and the motivations for our proposed system. 
\begin{figure}[htbp]
\centering
\includegraphics[width=1.0\linewidth]{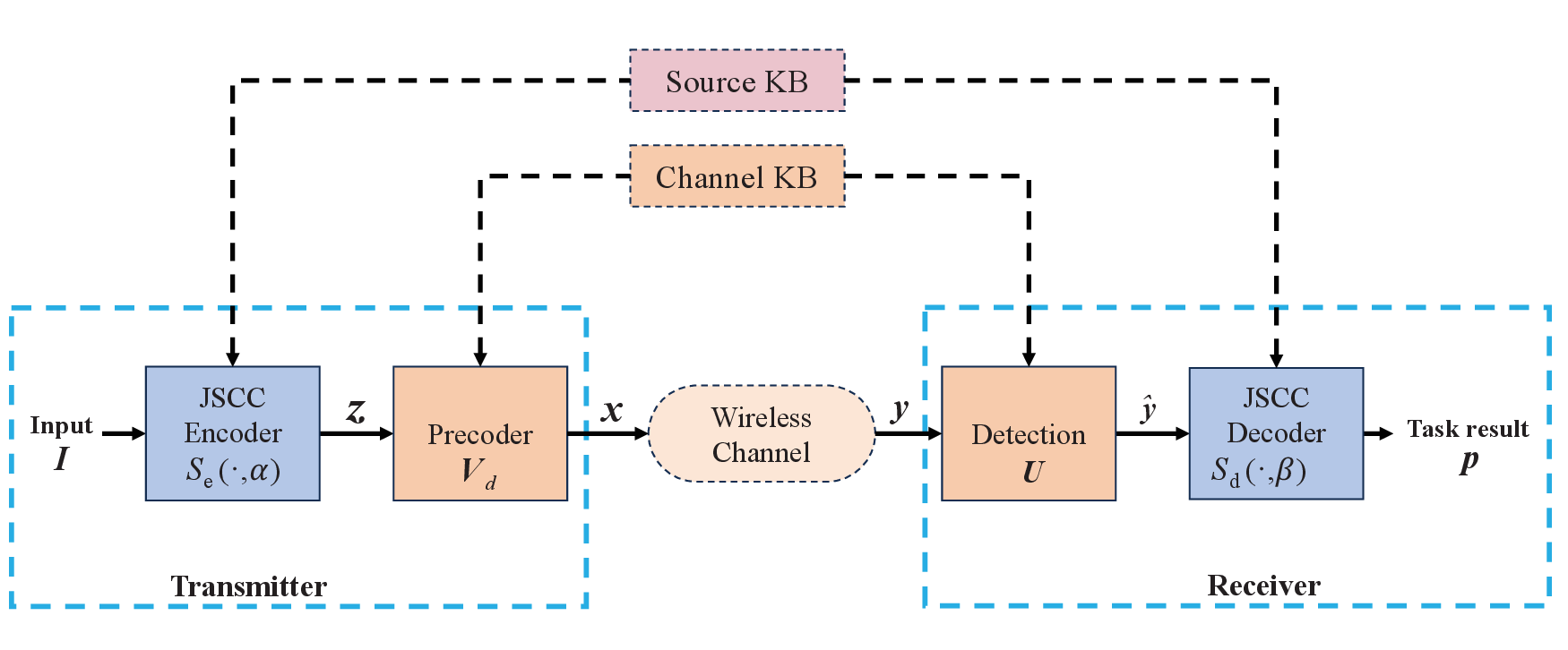}
\caption{SC system with a deep JSCC architecture.}
\label{fig1}
\end{figure}

\subsection{System Model}
Fig. \ref{fig1} shows the considered SC system with deep joint source and channel coding (JSCC) architecture. In particular, the transmitter consists of a JSCC encoder to extract semantic features and a precoder module to facilitate wireless transmission. The receiver is composed of a detection module for detecting the transmitted semantic features and a semantic decoder to accomplish underlying tasks. Moreover, a shared source KB that consists of task-related datasets is deployed to train the JSCC codec to perform underlying tasks jointly. Besides, a shared channel KB that consists of estimated CSI is deployed to enable precoding and detection to improve semantic transmission. 

In particular, at the transmitter, the source information $\bm{I}$ is first encoded by the JSCC encoder, which can be denoted by
\begin{equation}
\label{equ1}
\bm{z} = S_{\mathrm{e}}(\bm{I}, \bm{\alpha}),
\end{equation}
where \( S_{\mathrm{e}}(\cdot, \bm{\alpha})\) denotes the JSCC encoder with the parameter set \(\bm{\alpha}\) at the transmitter. Here, a MIMO channel is considered and assume the MIMO channel between the transmitter and receiver is denoted by \(\bm{H} \in \mathbb{C}^{N_\mathrm{r} \times N_\mathrm{t}}\), where \(N_\mathrm{t}\) and \(N_\mathrm{r}\) are the numbers of transmit and receive antennas, respectively. To enable efficient transmission, we perform singular value decomposition (SVD) on \(\bm{H}\) as
\begin{equation}
\bm{H} = \bm{U} \bm{\Sigma} \bm{V}^{\mathrm{H}},
\end{equation}
where \(\bm{U}\in\mathbb{C}^{N_\mathrm{r} \times N_\mathrm{r}}\) and \(\bm{V} \in \mathbb{C}^{N_\mathrm{t} \times N_\mathrm{t}}\) are unitary matrices, and \(\bm{\Sigma} \in \mathbb{C}^{N_\mathrm{r} \times N_\mathrm{t}}\) is a diagonal matrix whose non-zero elements represent the singular values of \(\bm{H}\).

The transmitter uses the first \(d\) right singular vectors \(\bm{V}_d \in \mathbb{C}^{N_\mathrm{t} \times d}\) to construct the precoding matrix.  To combat the channel noise, the extracted feature $\bm{z}$ is precoded as
\begin{equation}
\bm{x} = \bm{V}_d \bm{z}.
\end{equation}

As shown in Fig. \ref{fig1}, the semantic receiver contains the detection module and JSCC decoder to perform the underlying task. In particular, the received signal is given by
\begin{equation}
\bm{y} = \bm{H} \bm{x} + \bm{n} = \bm{U} \bm{\Sigma}_d \bm{z} + \bm{n},
\end{equation}
where \(\bm{\Sigma}_d\) is the truncated diagonal matrix and \(\bm{n} \sim \mathcal{CN}(0, \sigma^2 \bm{I})\) is the additive white Gaussian noise (AWGN). The receiver applies \(\bm{U}^{\mathrm{H}}\) to obtain
\begin{equation}
\hat{\bm{y}} = \bm{U}^{\mathrm{H}} \bm{y} = \bm{\Sigma}_d \bm{z} +  \bm{U}^{\mathrm{H}}\bm{n}.
\end{equation}

Then, the detection output is sent to the JSCC decoder to obtain the task result $\bm{p}$, which is denoted by
\begin{equation}
\bm{p}= S_{\mathrm{d}}(\hat{\bm{y}}, \bm{\beta}),
\end{equation}
where $S_{\mathrm{d}}(\cdot, \bm{\beta})$ denotes the JSCC decoder with the parameter set $\bm{\beta}$. 
Let $\mathcal{F}(\bm{I}; \bm{\alpha}, \bm{\beta}, \bm{H})$ denote the end-to-end mapping from the input $\bm{I}$ to the task output $\bm{p}$, encompassing the semantic encoder, MIMO channel with precoding and detection, and the semantic decoder. Then, the optimization objective is
\begin{equation}
\label{ori_loss}
\min_{\bm{\alpha}, \bm{\beta}} \ \mathbb{E} \left[ \mathcal{L}_{\text{task}} \left( \mathcal{F}(\bm{I}; \bm{\alpha}, \bm{\beta}, \bm{H}), \hat{\bm{p}} \right) \right],
\end{equation}
where \(\mathcal{L}_{\text{task}}\) denote the loss function measuring the discrepancy between the task result $\bm{p}$ and the ground truth $\hat{\bm{p}}$. The mapping $\mathcal{F}$ is defined as
\begin{equation}
\mathcal{F}(\bm{I}; \bm{\alpha}, \bm{\beta}, \bm{H}) = S_d\left( \bm{U}^\mathrm{H} \left( \bm{H} \bm{V}_d S_e(\bm{I}, \bm{\alpha}) + \bm{n} \right), \bm{\beta} \right).
\end{equation}

\begin{figure}[t!]
\centering
\includegraphics[width=1.0\linewidth]{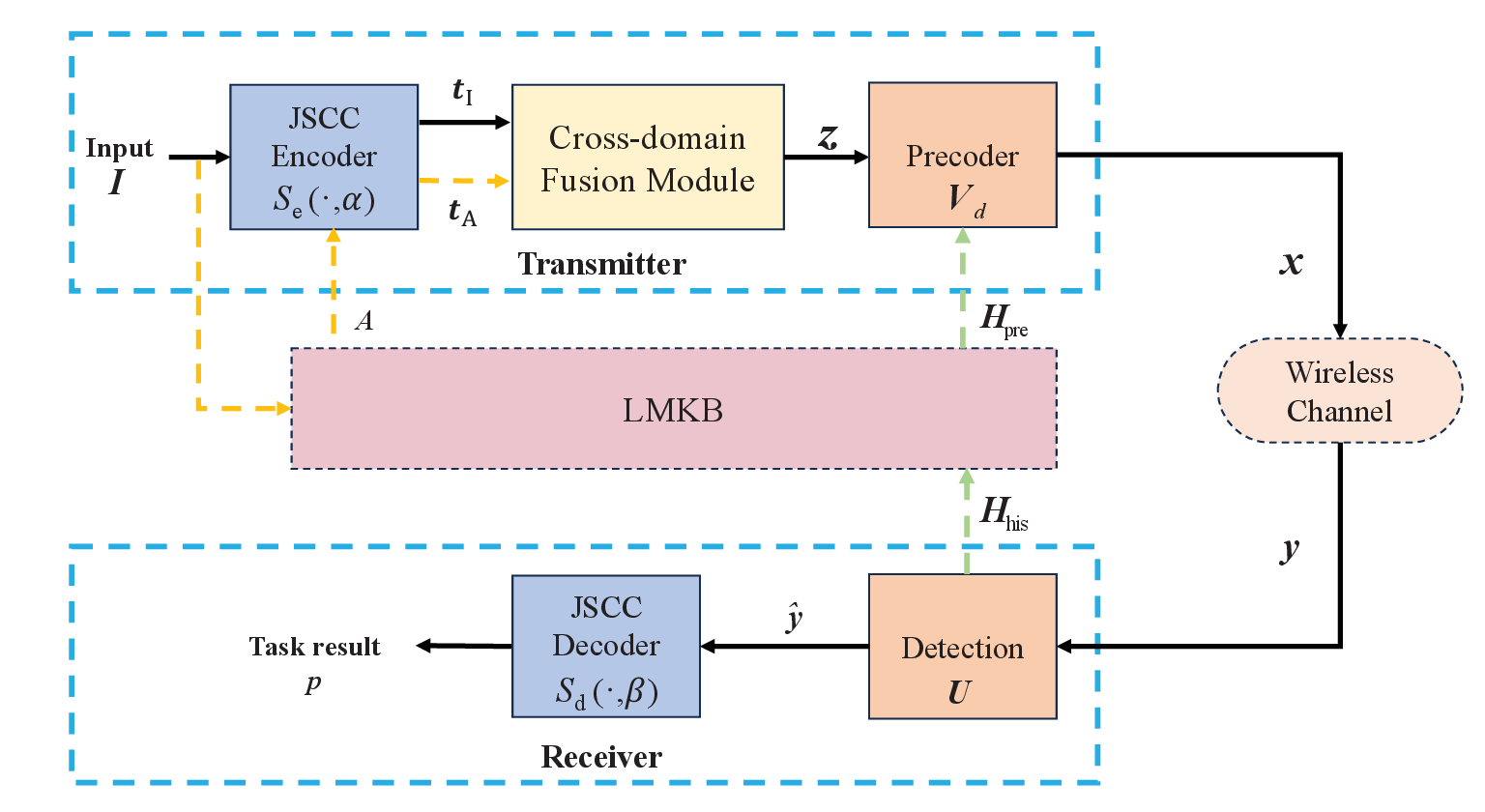}
\caption{Illustration of the proposed SC-LMKB.}
\label{fig2}
\end{figure}

\subsection{Motivation}
To minimize the task loss, the model requires both accurate CSI $\bm{H}$ and sufficient source data \(\bm{I}\). We define the source KB and channel KB as a training dataset containing \(N\) samples, which can be expressed as
\begin{equation}
\mathcal{D}_{\text{train}} = \left\{ (\bm{I}_i, \bm{H}_i) \right\}_{i=1}^{N},
\end{equation}
where each tuple includes a source input $\bm{I}_i$, its corresponding CSI $\bm{H}_i$. With sufficient data in the semantic KBs, the above objective allows the model to generalize well on unseen data. However, most existing works construct KB directly from task-specific datasets,  which are often limited in scale and diversity\cite{liukg}. Besides, acquiring large-scale and diverse datasets is challenging in practical SC scenarios, such as IoT. Specifically, the source data $\bm{I}$ often suffers from sparse annotations, making it difficult to provide comprehensive semantic diversity. For channel KB, the conventional pilot-based method sends pilots to the receiver. Then, the receiver estimates the CSI and feeds it back to the transmitter, which incurs additional transmission overhead. 
To address the above limitations, we propose incorporating an LLM as a KB to generate data for both source KBs and channel KBs. Specifically, we denote the generated dataset as
\begin{equation}
\mathcal{D}_{\text{aug}} = \mathcal{D}_{\text{train}} \cup \left\{ (\tilde{\bm{I}}_j, \tilde{\bm{H}}_j) \right\}_{j=1}^{M},
\end{equation}
where \(M\) is the size of the generated dataset. The generated semantic–channel pairs \((\tilde{\bm{I}}_j, \tilde{\bm{H}}_j)\) are produced by an LMKB, denoted as 
\begin{equation}
\label{constrain1}
(\tilde{\bm{I}}_j, \tilde{\bm{H}}_j) = f_{\text{LLM}}(\bm{I}_{j}, \bm{H}_{j} ).
\end{equation}

Although LLMs have demonstrated powerful generative capabilities, hallucinations in the generated data may introduce semantic noise and degrade task performance \cite{hallucination}. In particular, hallucinated data can break the semantic consistency between the source data and its corresponding task labels, leading to misleading supervision during training.
Moreover, hallucinations may cause the generated channel data to deviate from the true distribution, resulting in mismatches between the designed precoding matrix \(\bm{V}_{d}\) and the actual channel conditions. To address the challenge, we constrain the generated data to remain close to the original data. Specifically, we filter out the generated pair \((\tilde{\bm{I}}_j, \tilde{\bm{H}}_j)\) that exceeds a predefined threshold \(\epsilon\), which can be expressed as
\begin{equation}
\label{constrain2}
\text{Dist}\left( (\tilde{\bm{I}}_j, \tilde{\bm{H}}_j), (\bm{I}_j, \bm{H}_j) \right) < \epsilon.
\end{equation}

Combining (\ref{constrain1}) and (\ref{constrain2}), the optimization objective (\ref{ori_loss}) can be reformulated by solving
\begin{align}
\label{new_equation}
\min_{\bm{\alpha}, \bm{\beta}} \quad & \mathbb{E}_{(\bm{I}, \bm{H}, \bm{y}) \sim \mathcal{D}_{\text{aug}}} \left[ \mathcal{L}_{\text{task}}\left( \mathcal{F}(\bm{I}; \bm{\alpha}, \bm{\beta}, \bm{H}), \hat{\bm{p}} \right) \right] \\
\text{s.t.} \quad & (\tilde{\bm{I}}_j, \tilde{\bm{H}}_j) = f_{\text{LLM}}(\bm{I}_j, \bm{H}_j), \quad \forall j \in \{1, \dots, M\} \\
& \text{Dist}\left( (\tilde{\bm{I}}_j, \tilde{\bm{H}}_j), (\bm{I}_j, \bm{H}_j) \right) < \epsilon, \quad \forall j,
\end{align}
where the first constraint corresponds to \textit{Challenge 1}, ensuring that the LMKB generates additional data. The second constraint addresses \textit{Challenge 2} by filtering out hallucinated data whose source or channel content deviates significantly from the original data.

To solve the problem (\ref{new_equation}), we propose SC-LMKB, which leverages an LMKB to provide additional data for both source KB and channel KB, thereby empowering the JSCC codec to perform the downstream task more accurately. The detailed architecture and implementation of SC-LMKB are presented in Section~\ref{SC-LMKB}.

\section{SC-LMKB}
\label{SC-LMKB}
This section first introduces the overall architecture of SC-LMKB. Then, an LLM-enabled data generation mechanism is proposed, which leverages the LMKB for both source and channel data generation. In particular, a prompt engineering strategy is proposed to accomplish the source data generation. For the channel data generation, we propose a cross-attention alignment method to align CSI features with the natural language modality in the LLM space.

\subsection{Overview of SC-LMKB}
Fig. \ref{fig2} illustrates the overall architecture of the proposed SC-LMKB framework, which introduces an LMKB to address the data scarcity issue in MIMO SC by generating multi-modal data, including both source and channel data.
At the transmitter, the LMKB first provides additional source data to enrich the training dataset. To alleviate the impact of hallucinated noise that may arise from LLM-generated samples, a CDFC framework is employed to filter out the hallucinated data and fuse the semantic representations from the original data domain and the generated data domain. Meanwhile, the LLM-generated channel data is used to assist in designing an accurate precoding matrix under the MIMO channel model.
At the receiver side, a detection module is applied to recover the transmitted semantic features, which are then decoded by a JSCC decoder to accomplish the downstream task.

\begin{figure*}[t]
\centering
\includegraphics[width= 0.8\linewidth]{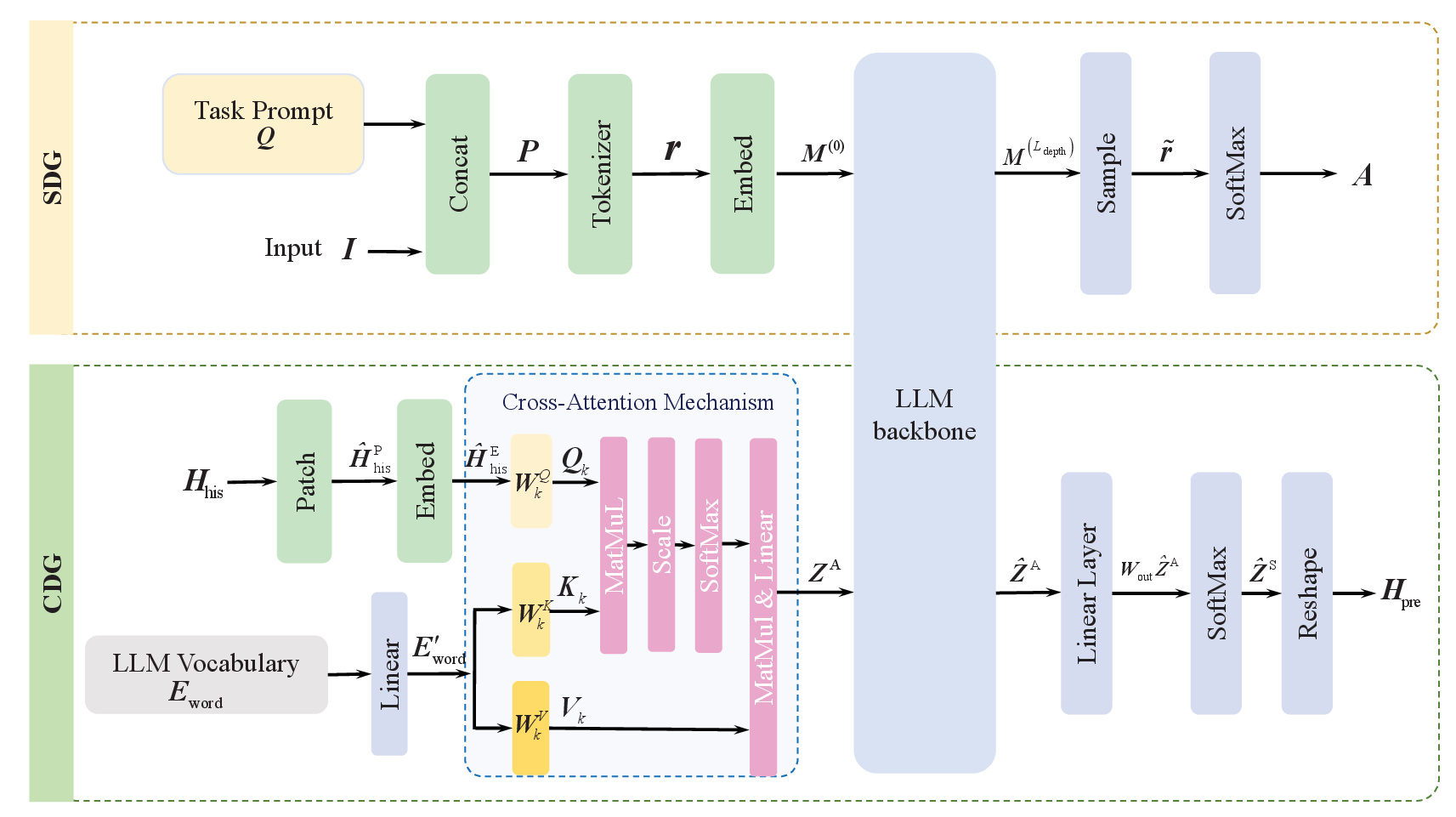}
\caption{Illustration of the proposed LMKB.}
\label{fig3}
\end{figure*}
\subsection{LLM-enabled Multi-modal Generation Mechanism}
To accomplish the LMKB, we propose an LLM-enabled multi-modal generation mechanism as shown in Fig. \ref{fig3}. Specifically, this mechanism integrates two modules: \textbf{(1) SDG module} and \textbf{(2) CDG module}. An LLM serves as the backbone to jointly process and generate multi-modal data.
\begin{figure}[htbp]
\centering
\includegraphics[width=0.8\linewidth]{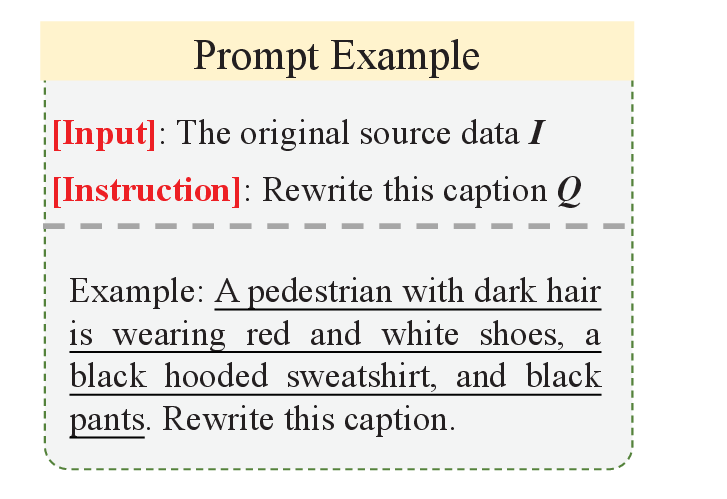}
\caption{A prompt example for SDG.}
\label{fig4}
\end{figure}

1) \textbf{SDG}: To support source data generation, we employ a prompt engineering strategy that guides a pre-trained language model to generate diverse and semantically aligned textual inputs. In particular, we first design a prompt according to the underlying task. Then, we tokenize and embed the concatenation of the prompt and original data. We input the concatenated embedding into the LLM backbone. Finally, we employed sampling to generate the output of the LLM.

As shown in Fig. \ref{fig3}, the strategy begins by designing a prompt that is aligned with the objectives of the task. In particular, the prompt consists of the original data \( \bm{I}\) and the instruction \( \bm{Q} \). The instruction provides specific guidance on how the LMKB should handle  \( \bm{I}\). For example, in a cross-model retrieval task, the instruction $\bm{Q}=\text{``Rewrite the caption''}$ as shown in Fig. \ref{fig4}. Mathematically, we express the task-specific prompt as $\bm{P} = [\bm{I; Q}]$. With the concatenated ${\bm{P}}$, a tokenizer \( \mathcal{V} \) is applied to convert the input sequence into discrete token indices according to a pre-trained vocabulary table, which can be expressed as
\[
\bm{r} = \mathcal{V}(\bm{P}) =[r_1, r_2, \dots, r_L],
\]
where \( \mathcal{V} \) denotes the tokenizer function that maps a string sequence to a sequence of vocabulary indices,  $L$ denotes the length of $\bm{r}$ and \( |\mathcal{V}| \) denotes the size of the vocabulary.

Then, the discrete token sequence \( \bm{r} \) is mapped into continuous embeddings via a token embedding matrix \( \bm{W}_\text{E} \in \mathbb{R}^{|\mathcal{V}| \times d_{\text{LLM}}} \). The resulting input embedding sequence $\bm{M}^{(0)}$ is expressed as
\[
\bm{M}^{(0)} = \bm{W}_\text{E}[\bm{r}] \in \mathbb{R}^{L \times d_{\text{LLM}}},
\]
where \( d_{\text{LLM}} \) denotes the hidden dimension of the language model. For example, GPT-2 uses \( d_{\text{LLM}} = 768 \), while LLaMA-7B uses \( d_{\text{LLM}} = 4096 \).

The embedding sequence $\bm{M}^{(0)}$ is then processed by the LLM backbone. With LLM backbone \( f_{\text{LLM}} \), the output hidden state $\bm{M}^{(L_{\text{depth}})}$ are given by
\begin{align}
\bm{M}^{(L_{\text{depth}})} 
&= f_{\text{LLM}}(\bm{M}^{(0)}) \nonumber \\
&= [\bm{m}^{(L_{\text{depth}})}_{1}, \bm{m}^{(L_{\text{depth}})}_{2}, \dots, \bm{m}^{(L_{\text{depth}})}_{L} ] 
\in \mathbb{R}^{L \times d_{\text{LLM}}},
\end{align}
where \( f_{\text{LLM}} \) consists of $L_{\text{depth}}$ transformer layers. Finally, the LLM backbone generates an output sequence \( \tilde{\bm{r}} \) by sampling from the conditional distribution defined over the final hidden states $\bm{M}^{(L_{\text{depth}})}$. Using a decoding strategy such as temperature scaling~\cite{renze2024effect}, the temperature-controlled sampling first rescales the predicted logits $\bm{z}_t$ by a temperature parameter $\tau > 0$, yielding a softened distribution:
\[
p(\tilde{\bm{r}}) = \text{softmax}\left(\frac{\bm{M}^{(L_{\text{depth}})}}{\tau}\right).
\]

The decoded output \( \tilde{\bm{r}} \) is then parsed to construct the generated source data \( \bm{A} \). The property of sampling from a probability distribution arises hallucination that may generate source data $\bm{A}$ with a different semantics compared with the original source data $\bm{I}$. Thus, a filtering mechanism is proposed in section \ref{cdfc}. \\

2) \textbf{CDG}: To enable CDG, we propose a cross-attention alignment method to generate predicted future CSI based on historical CSI.
Specifically, we first pre-process CSI features into a discrete token representation compatible with the LLM input space based on a cross-attention mechanism. 
Then, the aligned CSI tokens are input to the same LLM backbone as the SDG for token prediction. Finally, we post-process the predicted CSI tokens to convert tokens into CSI.

As illustrated in Fig.~\ref{fig3}, the generation process begins with the preprocessing of historical CSI to make it compatible with the LLM input format. Let the historical CSI and predict CSI be denoted as \( \bm{H}_\text{his} \in \mathbb{C}^{T_{\text{his}} \times N_\text{r} \times N_\text{t}} \) and \( \bm{H}_\text{pre} \in \mathbb{C}^{T_{\text{pre}} \times N_\text{r} \times N_\text{t}} \), respectively. \( T_\text{his} \) and \( T_\text{pre} \) is the number of history and predict time. 
Since the LLM can only handle real numbers, we arrange the complex \( \bm{H}_\text{his} \) into real tensors \( \bm{H}_\text{his}^{\text{R}} \in \mathbb{R}^{T_{\text{his}} \times N_\text{r} \times N_\text{t} \times 2} \).

To ensure stable training, we apply normalization to \( \bm{H}_\text{his}^{\text{R}} \). The normalized CSI is computed as
\[
\hat{\bm{H}}_\text{his}^{\text{R}}= \frac{\bm{H}_\text{his}^{\text{R}} - \mu_{\text{his}}}{\sigma_{\text{his}}},
\]
where \( \mu_{\text{his}} \) and \( \sigma_{\text{his}} \) denote the mean and standard deviation computed over all entries in \( \bm{H}_\text{his}^{\text{R}} \).

To capture local temporal structures, we apply a sliding window patching operation to segment the normalized CSI sequence ${\hat{\bm{H}}_\text{his}^{\text{R}}}$ into overlapping patches. Specifically, the normalized CSI ${\hat{\bm{H}}_\text{his}^{\text{R}}}$ is partitioned into patches ${\hat{\bm{H}}_\text{his}^{\text{P}} \in \mathbb{R}^{2 N_\text{r}N_\text{t} \times N_{\text{patch}} \times L_{\text{patch}} } }$ 
where ${N_{\text{patch}}=\frac{T_{\text{his}}-L_{\text{patch}}}{S}}$ denotes the number of patches. $S$ denotes the sliding stride, and ${L_{\text{patch}}}$ denotes the length of patches. 

To transform each CSI patch into a latent representation, CSI embedding module is employed to map ${\hat{\bm{H}}_\text{his}^{\text{P}} \in \mathbb{R}^{2 N_\text{r}N_\text{t} \times N_{\text{patch}} \times L_{\text{patch}} } }$ into an embedding vector ${\hat{\bm{H}}_\text{his}^{\text{E}} \in \mathbb{R}^{2 N_\text{r}N_\text{t} \times N_{\text{patch}} \times d_{\text{E}} } }$ where $d_{\text{E}}$ denotes the embedding dimension.
The resulting CSI embeddings ${\hat{\bm{H}}_\text{his}^{\text{E}}}$ are treated as the query input to a cross-attention module.

Inspired by\cite{jin2023time}, we adopt a cross-modal attention mechanism that aligns the CSI embedding with the pretrained LLM word embedding space. For clarity, we denote the CSI embedding of the $i$-th sample as  ${\hat{\bm{H}}_\text{his}^{\text{E}~(i)}} \in \mathbb{R}^{N_{\text{patch}} \times d_{\text{E}}}$.
Specifically, let the LLM word embedding matrix be denoted as \(\bm{E}_{\text{word}} \in \mathbb{R}^{|\mathcal{V}| \times d_\text{LLM}} \). For example, GPT-2 has $|\mathcal{V}|= 50257$, while LLaMA-7B has $|\mathcal{V}|=32000$.
Due to the large size of ${\bm{E}_{\text{word}}}$, it is computationally expensive to directly use \( E_{\text{word}} \) as the key and value inputs in the attention mechanism. Thus, a learnable projection matrix \( \bm{W}_{\text{proj}} \in \mathbb{R}^{d_\text{LLM} \times \hat{d}_{\text{LLM}}}\) is introduced to reduce the dimensionality of the word embeddings. The compressed LLM word ${\bm{E}_{\text{word}}^{'}}$ can be expressed as
\begin{equation}{\bm{E}_{\text{word}}^{'}} = {\bm{E}_{\text{word}}}\bm{W}_{\text{proj}}.
\end{equation}

\begin{figure*}[htbp]
\centering
\includegraphics[width=0.80\linewidth]{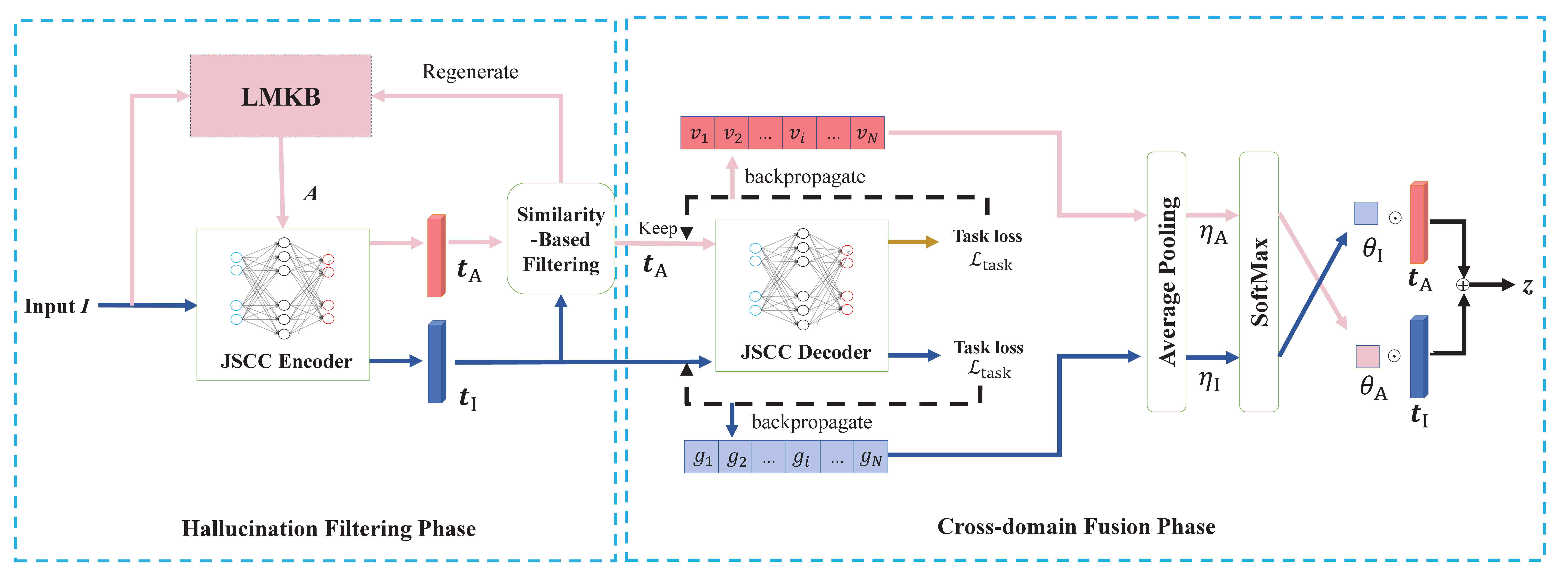}
\caption{A cross-domain fusion codec framework.}
\label{fig5}
\end{figure*}
Subsequently, a multi-head cross-attention layer with $k$ heads can be employed, which can be expressed as
\begin{equation}
    \bm{Z}_{k}^{\text{A}(i)} = \mathrm{Softmax} \left( \frac{\bm{Q}_{k}^{(i)} \cdot (\bm{K}_{k}^{(i)})^\top}{\sqrt{d_{k}}} \right) \cdot \bm{V}_{k}^{(i)},
\end{equation}
where the query matrix \( \bm{Q}_{k}^{(i)}= {\hat{\bm{H}}_\text{his}^{\text{E}~(i)}}\bm{W}_{k}^{Q}\) , the key and value matrices are computed as \( \bm{K}_{k}^{(i)} = {\bm{E}_{\text{word}}^{'}} \bm{W}_{k}^{K} \) and \( \bm{V}_{k}^{(i)} = {\bm{E}_{\text{word}}^{'}} \bm{W}_{k}^{V} \), respectively. This operation enables the CSI representation to selectively attend to relevant tokens in the LLM word embeddings ${\bm{E}_{\text{word}}^{'}}$. Aggregating $\bm{Z}_{k}^{\text{A}(i)}$ in each head, we obtain $\bm{Z}^{\text{A}(i)}\in \mathbb{R}^{N_{\text{patch}}\times d_{\text{E}}}$.
After applying the cross-attention alignment mechanism, the CSI patches are projected into the token embedding space, forming a sequence \( \bm{Z}^{\text{A}(i)} \), where each row corresponds to an aligned embedding of a CSI patch.

With the cross-attention mechanism, the CSI patches are projected into the token embedding space and can be seen as a series of ``sentences''. This structure enables us to treat the channel modeling task as a sequential prediction problem, similar to next-token prediction in LLM. Specifically, $\bm{Z}^{\text{A}(i)} = [\bm{z}_1^{\text{A}}, \bm{z}_2^{\text{A}}, \dots, \bm{z}_{N_{\text{patch}}}^{\text{A}}]$ is processed by the LLM backbone, which can be expressed as
\begin{equation}
    \hat{\bm{Z}}^{\text{A}(i)} = f_{\text{LLM}}(\bm{Z}^{\text{A}(i)}) = [\hat{\bm{z}}_1^{\text{A}}, \hat{\bm{z}}_2^{\text{A}}, \dots, \hat{\bm{z}}_{N_{\text{patch}}}^{\text{A}}].
\end{equation}

To evaluate the prediction performance, we apply a linear output layer ${\bm{W}_{\text{out}}}$ followed by a softmax function to produce the probability distribution over the vocabulary, which can be denoted by
\begin{equation}
\label{ZNS}
    \hat{\bm{z}}_n^{\text{S}}=\text{softmax}(\bm{W}_{\text{out}} \cdot \hat{\bm{z}}_n^{\text{A}})
\end{equation}

Aggerating $\hat{\bm{z}}_n^{\text{S}}$, we can obtain $\hat{\bm{Z}}^{\text{S}}$. To map the predicted $\hat{\bm{Z}}^{\text{S}}$ back to the CSI domain, a linear projection layer \( \bm{W}_{\text{linear}} \in \mathbb{R}^{N_\mathrm{r} N_\mathrm{t} \times T_{\text{pre}}} \) is applied to each predicted embedding. Thus, we can obtain the predict CSI $\hat{\bm{H}}_{\text{pre}}$, which can be denoted by
\begin{equation}
\hat{\bm{H}}_{\text{pre}} = \text{Reshape}(\hat{\bm{Z}}^{\text{S}} \bm{W}_{\text{linear}}) \in \mathbb{C}^{T_{\text{pre}} \times N_\mathrm{r} \times N_\mathrm{t}}.
\end{equation}

Overall, with the proposed SDG and CDG modules, we can leverage a unified LLM backbone for data generation in both source and channel KBs without requiring additional fine-tuning LLM. However, the hallucination from LLMs may introduce semantic noise, thereby degrading the accuracy and relevance of the generated data. Thus, we introduce Section \ref{cdfc} to address the hallucination issue.

\section{Proposed hallucination mitigation method}
\label{cdfc}

In this section, we first introduce the overall architecture of the CDFC framework to alleviate the hallucination in SDG, which consists of a hallucination filtering phase and a cross-domain fusion phase. Then, a joint training objective is proposed to reduce the impact of hallucination on CDG.
Finally, the training and inference stage of SC-LMKB is proposed. 

\subsection{CDFC}
To resist the semantic noise caused by hallucination from LLMs, a CDFC framework is proposed as shown in Fig. \ref{fig5}. In particular, the framework consists of a hallucination filtering phase and a cross-domain fusion phase.
\subsubsection{Hallucination Filtering Phase}
The hallucination filtering phase is designed to filter out source data generated by LMKB that deviates from the intended context. Specifically, the source information $\bm{I}$ is encoded by the JSCC encoder, which can be denoted by
\begin{equation}
\bm{t}_{\mathrm{I}} = S_{\mathrm{e}}(\bm{I}, \bm{\alpha}),
\label{encoder}
\end{equation}
where \( S_{\mathrm{e}}(\cdot, \bm{\alpha})\) denotes the JSCC encoder with the parameter set \(\bm{\alpha}\). Meanwhile, the source information $\bm{I}$ is inputted into the LMKB with a task-related prompt $\bm{P}$. Let the output from the LMKB be denoted by $\bm{A}$. Then, $\bm{A}$ is encoded into the semantic feature $\bm{t}_{\mathrm{A}}$ using the same JSCC encoder as in (\ref{encoder}), which can be expressed as
\begin{equation}
\bm{t}_{\mathrm{A}} = S_{\mathrm{e}}(\bm{A}, \bm{\alpha}).
\end{equation}\par
Due to the hallucination of LLMs, the generated data from the LMKB $\bm{A}$ may not align with the source $\bm{I}$, potentially resulting in inconsistent or even incorrect task outcomes. Thus, a similarity-based filtering algorithm is applied to remove data generated by LMKB that deviates from the intended context\cite{li2024data}. Specifically, the relevance between the semantic features $\bm{t}_{\mathrm{I}}$ and $\bm{t}_{\mathrm{A}}$ is quantified by the cosine similarity
\begin{equation}
\label{similarity}
\text{sim}(\bm{t}_{\mathrm{I}}, \bm{t}_{\mathrm{A}}) = \frac{\bm{t}_{\mathrm{I}} \cdot \bm{t}_{\mathrm{A}}}{\|\bm{t}_{\mathrm{I}}\| \|\bm{t}_{\mathrm{A}}\|},
\end{equation}
where \( \cdot \) denotes the dot product and \(\|\cdot\|\) represents the Euclidean norm. A threshold \( \gamma \) is then defined to filter out hallucinated content. If the similarity score $\text{sim}(\bm{t}_{\mathrm{I}}, \bm{t}_{\mathrm{A}})$ exceeds this threshold \( \gamma \), the output of the LMKB $\bm{t}_{\mathrm{A}}$ is considered semantic relevant to $\bm{t}_{\mathrm{I}}$, and thus $\bm{t}_{\mathrm{A}}$ is also inputted to the cross-domain fusion module. Otherwise,  $\bm{t}_{\mathrm{A}}$ is discarded and the LMKB regenerates a new output until the similarity is above the threshold.
This process is outlined in \textbf{Algorithm 1}.\par
\begin{algorithm}[t!]
\label{algo}
\caption{Similarity-Based Filtering Algorithm}
\begin{algorithmic}[1]
\STATE Input: $\bm{t}_{\mathrm{I}}$, $\bm{t}_{\mathrm{A}}$.
\STATE Compute the similarity score $\text{sim}(\bm{t}_{\mathrm{I}}, \bm{t}_{\mathrm{A}})$ according to (\ref{similarity}).
\IF{$\text{sim}(\bm{t}_{\mathrm{I}}, \bm{t}_{\mathrm{A}}) > \gamma$}
    \STATE Keep $\bm{t}_{\mathrm{A}}$.
\ELSE
    \STATE Regenerate $\bm{t}_{\mathrm{A}}$ using the LMKB.
\ENDIF
\end{algorithmic}
\end{algorithm}
\subsubsection{Cross-domain Fusion Phase}
By implementing a similarity-based filtering algorithm, we filter out the data generated by LMKB that are considered irrelevant to the source data. However, there still exists noise that is irrelevant to the underlying task. Thus, a cross-domain fusion phase is proposed to select the most relevant semantic features with semantic importance to reduce task-irrelevant noise. To achieve this, it is crucial to evaluate the semantic importance of task performance. The core insight is that data with higher semantic importance tends to exhibit less semantic noise. With insights from Grad-CAM\cite{selvaraju2017grad}, we use feature gradients obtained through back-propagation with respect to the task loss $\mathcal{L}_{\text{task}}$ to quantify the significance of different semantic features. Specifically, we input $\bm{t}_{\mathrm{I}}$ and $\bm{t}_{\mathrm{A}}$ into the JSCC decoder to calculate the task loss through forward propagation. The gradients of the task loss with respect to each feature can then be computed through back-propagation.
For the source feature $\bm{t}_{\mathrm{I}}$, the gradients are expressed as
\begin{equation}
[g_1, g_2, \dots, g_N] = \left[ \frac{\partial \mathcal{L}_{\text{task}}}{\partial \bm{t}_{\mathrm{I}, 1}}, \frac{\partial \mathcal{L}_{\text{task}}}{\partial \bm{t}_{\mathrm{I}, 2}}, \dots, \frac{\partial \mathcal{L}_{\text{task}}}{\partial \bm{t}_{\mathrm{I}, N}} \right],
\end{equation}
where \( N \) is the feature dimension as defined by the JSCC encoder.
Similarly, for the additional feature  $\bm{t}_{\mathrm{A}}$ provided by LMKB, the gradients are represented by
\begin{equation}
[v_1, v_2, \dots, v_N] = \left[ \frac{\partial \mathcal{L}_{\text{task}}}{\partial \bm{t}_{\mathrm{A}, 1}}, \frac{\partial \mathcal{L}_{\text{task}}}{\partial \bm{t}_{\mathrm{A}, 2}}, \dots, \frac{\partial \mathcal{L}_{\text{task}}}{\partial \bm{t}_{\mathrm{A}, N}} \right],
\end{equation}\par
Then, these gradients are subsequently globally average-pooled to yield feature importance weights, which can be denoted by
\begin{equation} 
\eta_{\text{I}} = \frac{1}{N} \sum_{i=1}^{N} \frac{\partial \mathcal{L}_{\text{task}} }{\partial \bm{t}_{\mathrm{I}, i}}, \quad \eta_{\text{A}} = \frac{1}{N} \sum_{i=1}^{N} \frac{\partial \mathcal{L}_{\text{task}}}{\partial \bm{t}_{\mathrm{A}, i}}.
\end{equation}\par
To balance these pooled gradients, we apply a softmax layer, which can be expressed as
\begin{equation} 
\theta_{\text{I}} = \frac{e^{\eta_{\text{I}}}}{e^{\eta_{\text{I}}} + e^{\eta_{\text{A}}}}, \theta_{\text{A}} = \frac{e^{\eta_{\text{A}}}}{e^{\eta_{\text{I}}} + e^{\eta_{\text{A}}}}.
\end{equation}\par
To enable the JSCC codec to learn more effective knowledge from LMKB, the gradients are cross-multiplied with their corresponding features to obtain the final feature representation, as shown in Fig. \ref{fig3}. Specifically, the cross-multiplication can be expressed as follows:
\begin{equation}
\bm{z} = \theta_{\text{A}} \cdot \bm{t}_{\mathrm{I}} + \theta_{\text{I}} \cdot \bm{t}_{\mathrm{A}},
\end{equation}
where $\theta_{\text{A}}$ and $\theta_{\text{I}} $ are the weights obtained from the softmax layer based on gradient pooling, as defined previously. This cross-multiplication allows each domain's feature to be weighted by the gradient of the other domain, enabling the JSCC codec to learn the knowledge most relevant to the task while reducing noise from less informative features. By emphasizing task-relevant information and minimizing irrelevant or redundant information, this approach can resist the semantic noise introduced by the hallucination problem and improve task accuracy.

\subsection{Joint training objective}
To reduce the impact of hallucination on CDG, we propose a joint training objective that combines cross-entropy loss and reconstruction loss. The joint training objective enables the model to simultaneously learn token-level semantics and signal-level structures for accurate channel-related data generation.

With the cross-attention mechanism in Section \ref{SC-LMKB}, the historical CSI ${}$ can be converted to LLM token embeddings. With the predicted tokens $\hat{\bm{z}}_n^{\text{S}}$ in (\ref{ZNS}), the cross-entropy loss is given by
\begin{equation}
\label{CEloss}
    \mathcal{L}_{\text{CE}} = - \sum_{n=1}^{{N_{\text{patch}}}-1} \bm{p}_{n+1}\log (\hat{\bm{z}}_n^{\text{S}}),
\end{equation}
where \( \bm{p}_{n+1} \in \mathbb{R}^{|\mathcal{V}|} \) denotes the one-hot encoded vector corresponding to the ground-truth token at step \( n+1 \) and $\hat{\bm{z}}_n^{\text{S}}=\text{softmax}(\bm{W}_{\text{out}} \cdot \hat{\bm{z}}_n^{\text{A}})$ denotes the predicted token probability distribution at step \( n \).

Since the CE loss (\ref{CEloss}) ensures the token-level precision by maximizing the probability of generating correct tokens, it supervises the model to produce a token sequence that closely matches the ground truth. However, the hallucination from LLM may predict fake tokens that corrupt the reconstructed CSI and lead to a mismatch between the generated and actual CSI. Therefore, a reconstruction loss based on the normalized mean squared error (NMSE) between the predicted CSI $\hat{\bm{H}}_{\text{pre}}$ and the ground-truth CSI $\bm{H}_{\text{pre}}$ is adopted to ensure the signal-level precision, which can be expressed by
\begin{equation}
\mathcal{L}_{\text{NMSE}} = \frac{\| \bm{\hat{H}}_{\text{pre}} - \bm{H}_{\text{pre}} \|_F^2}{\| \bm{H}_{\text{pre}} \|_F^2},
\end{equation}
where  $ \| \cdot \|_F^2$ represents the squared Frobenius norm.
The final loss combines both the token-level cross-entropy loss and the NMSE reconstruction loss to ensure consistency in the latent and CSI domains
\begin{equation}
\mathcal{L}_{\text{total}} = \mathcal{L}_{\text{CE}} + \lambda \mathcal{L}_{\text{NMSE}}
\end{equation}
where \( \lambda \) is a weighting factor controlling the trade-off between sequence modelling and CSI accuracy. With the combined loss, we can prevent the generated channel data of the LMKB from hallucination. 

\subsection{Training and Inference stage}
The training stage of the proposed SC-LMKB aims at utilizing the fused $\bm{z}$ in the cross-domain fusion codec framework to train the whole JSCC codec. In particular, $\bm{z}$ is transmitted to the JSCC decoder and obtains the task result $\bm{p}$. The task loss is calculated by the $\mathcal{L}_{\text{task}}$ and updates the parameters of the JSCC codec by stochastic gradient descent (SGD). At the inference stage, the transmitter directly transmits the source features extracted over a wireless MIMO channel to accomplish downstream tasks at the receiver by the pre-trained JSCC codec.

\section{Experiments}
In this section, we demonstrate the superiority of the proposed SC-LMKB by numerical results.

\subsection{Experiment Setup}
\textit{1) Dataset:} In this work, we focus on the text-based cross-modal retrieval task, which is known for the challenges in obtaining high-quality datasets due to the high costs of annotation \cite{li2024data}. 
In particular, text-based person retrieval (TPR), text-based audio retrieval (TAR) and text-based motion retrieval (TMR) are considered.
\begin{itemize}
    \item \textbf{TPR Dataset}: The TPR task refers to transmitting the text description to retrieve the most relevant pedestrian image at the receiver. For performance evaluation, we have conducted SC-LMKB on TRP across three datasets, including CUHK-PEDES\cite{CUHK}, ICFG-PEDES\cite{ICFG}, and RSTPReid\cite{RSTP}. 
    %CUHK-PEDES is the first collection designed specifically for TPR, including 40,206 images and 80,412 textual descriptions corresponding to 13,003 unique identities. Besides, the ICFG-PEDES dataset has 54,522 images linked to 4,102 identities, with each image accompanied by a single textual description. Furthermore, the RSTPReid dataset contains 20,505 images of 4,101 identities captured across 15 cameras. 
    All datasets are divided into training and testing sets according to the official benchmark protocols.
    
    \item \textbf{TMR Dataset}: The TMR task refers to transmitting a textual description to retrieve the most semantically relevant motion sequence at the receiver, which is evaluated on the KIT Motion-Language dataset. This dataset consists of 3,911 full-body motion recordings represented in the Master Motor Map format, each accompanied by one or more textual descriptions. In total, there are 6,278 natural language annotations in English, describing various human actions. 
        
    \item \textbf{TAR Dataset}: The TAR task refers to transmitting a textual description to retrieve the most semantically relevant audio clip at the receiver. The TAR task is conducted on the Clotho v2 dataset \cite{drossos2020clotho}, which comprises 3,839 audio clips in the training set and 1,045 audio clips each in the validation and test sets. Each audio clip is annotated with five diverse human-written captions, with lengths ranging from 8 to 20 words, covering a wide variety of everyday acoustic scenes.
\end{itemize}

\textit{2) Evaluation Metrics:} For a comprehension evaluation, we adopt mean average precision (mAP) as the evaluation metric for the TPR task. mAP is a common metric used to evaluate the accuracy of information retrieval systems across an entire dataset. mAP evaluates overall ranking quality for multi-instance matching in TPR, whereas Rank@$k$ emphasizes top-$k$ retrieval success, which is more appropriate for TMR and TAR tasks. Thus, we adopt the popular Rank@$k$ (Rank@$k$ for short, $k$ = 1, 5, 10) as the evaluation metrics for TMR and TAR tasks. The higher Rank@$k$ and mAP indicate better performance. 

\textit{3) Model Deployment Details:} 
 For the TPR and TMR tasks, we utilize the CLIP backbone as the semantic codec to align image/motion representations with textual descriptions. For the TAR task, the BERT-base-uncased model is employed to encode textual queries. In our SC-LMKB framework, we adopt open-source LLMs, specifically Vicuna and LLaMA, as the backbone of the LMKB.
 For the MIMO channel, we adopt the widely used channel generator QuaDRiGa to simulate time-varying CSI datasets compliant with 3GPP TR 38.901. Specifically, we adopt the Urban Microcell (UMi) channel model under line-of-sight (LOS) conditions, denoted as UMi-LOS. This setup reflects short-range communication scenarios typical in dense urban deployments, such as street-level IoT nodes. In contrast, during the zero-shot generalization evaluation, we switch to the Urban Macrocell (UMa) model under non-line-of-sight (NLOS) conditions, denoted as UMa-NLOS. This simulates large-cell deployments where the signal experiences significant scattering, diffraction, and shadowing due to building obstructions and extended propagation paths. By testing on UMa-NLOS without fine-tuning, we assess the model's ability to generalize across different spatial scales and channel conditions, highlighting its robustness in unseen environments. Assume both the UMi-LOS and UMa-NLOS have the same MIMO antenna numbers as $N_\mathrm{t}$ = 16 and $N_\mathrm{r}$ = 16. All the experiments of SC-LMKB and other DL-based benchmarks are run on RTX4090 GPUs. 

\textit{4) Baselines:} For fair comparisons, we adopt the following baselines.
\begin{itemize}
    \item \textbf{DeepSC-MIMO}\cite{zhang2024scan}: DeepSC-MIMO extends the DeepSC framework to MIMO systems, where a deep learning-enabled SC model is deployed with pilot-based CSI estimation and feedback. This represents a typical feedback-based semantic transmission scheme under pilot-based CSI conditions.
    
    \item \textbf{LLM4CP}\cite{liu2024llm4cp}: LLM4CP leverages a LLM backbone to predict CSI through fine-tuning. Since the original design focuses solely on CSI generation, we integrate our proposed SDG module with LLM4CP to construct a complete end-to-end baseline.

    \item \textbf{Csi-LLM}\cite{fan2025csi}: Csi-LLM introduces a modality alignment mechanism that aligns CSI with the LLM token space to enable CSI prediction without feedback. Similar to LLM4CP, we extend this method by incorporating our SDG module to enable full pipeline comparisons.

    \item \textbf{SSCC}: As a classical separate source and channel coding scheme, this method applies Huffman coding for source compression and Reed-Solomon (RS) coding for channel protection. Quadrature Amplitude Modulation (QAM) is used for signal modulation, while SVD-based precoding and detection are adopted for the MIMO channel.
\end{itemize}
\begin{figure*}[htbp]
    \centering
    \subfloat[CUHK-PEDES]{
        \includegraphics[width=0.31\textwidth]{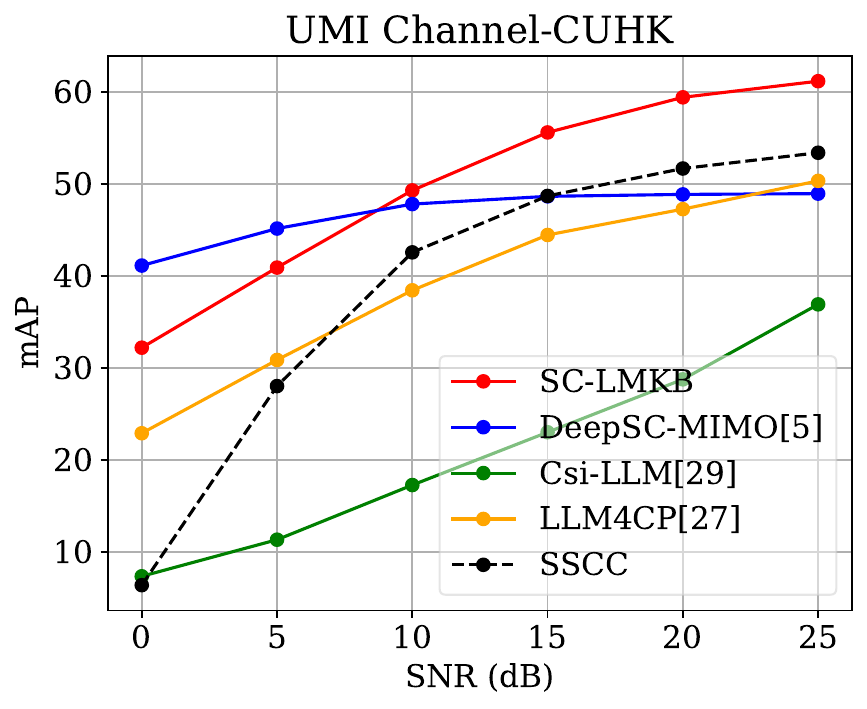}
        \label{figUMi:sub1}
    }
    \hfill
    \subfloat[ICFG-PEDES]{
        \includegraphics[width=0.31\textwidth]{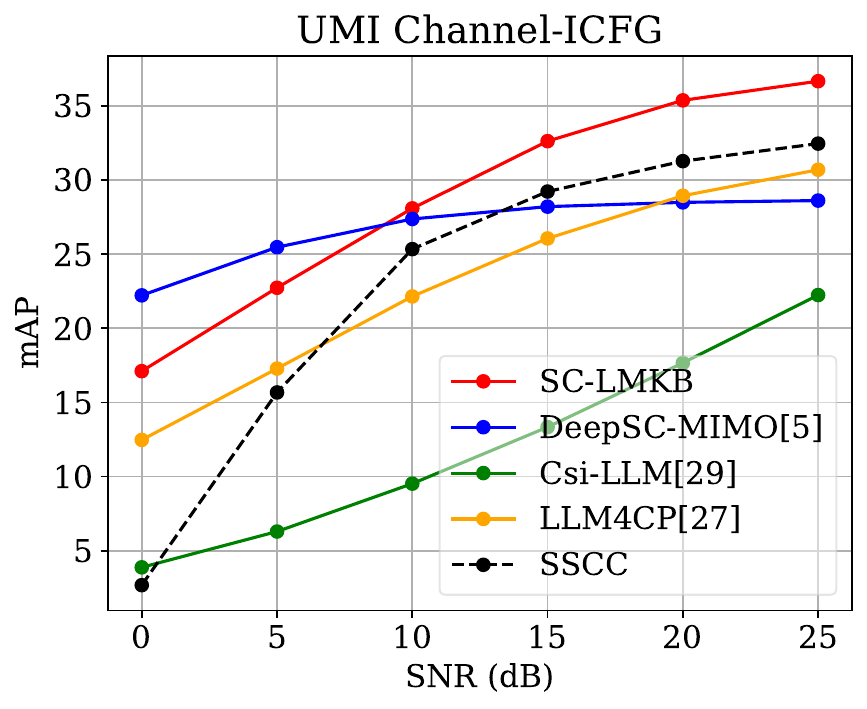}
        \label{figUMi:sub2}
    }
    \hfill
    \subfloat[RSTPReid]{
        \includegraphics[width=0.31\textwidth]{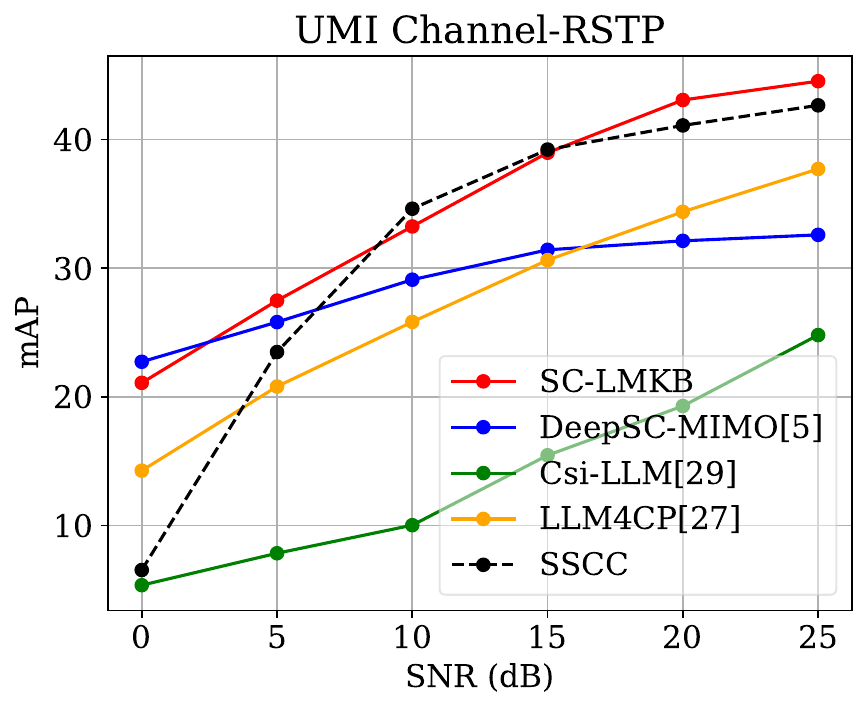}
        \label{figUMi:sub3}
    }
    \caption{mAP versus SNR over UMi-LOS channel on TPR datasets.}
    \label{figUMi}
\end{figure*}

\subsection{Performance Comparison}
To illustrate the effectiveness of the proposed SC-LMKB, we compare its performance with the baseline models under different SNR levels over the UMi-LOS channel. 

\textbf{Improvements on the TPR datasets.} Fig. \ref{figUMi} shows the mAP of all systems with SNRs over the UMi-LOS channel on the adopted TPR datasets. As shown in Fig.~\ref{figUMi}\subref{figUMi:sub1}, it can be observed that the mAP increases with SNR and gradually converges to a certain threshold. This is because higher SNR means a better communication condition and can decrease transmission errors, which consequently increases the task performance at the receiver. It can also be observed that the proposed SC-LMKB outperforms the LLM4CP, Csi-LLM and the traditional method across the entire SNR region. This demonstrates the effectiveness of our proposed LMKB. Besides, the proposed SC-LMKB method significantly outperforms DeepSC-MIMO, especially in high SNRs. Specifically, when SNR is 25 dB, SC-LMKB achieves improvements of 24.9\%, 65.6\%, and 21.5\% mAP gains compared to DeepSC-MIMO, Csi-LLM, and LLM4CP, respectively.
The results in Fig.~\ref{figUMi}\subref{figUMi:sub2} and Fig.~\ref{figUMi}\subref{figUMi:sub3} exhibit the similar trends in Fig.~\ref{figUMi}\subref{figUMi:sub1}. 
However, DeepSC-MIMO has a slightly better performance at low SNR regions as shown in Fig. \ref{figUMi} because low SNRs will affect the performance of the effect of CDG and CDFC, thus leading to performance degradation.
Besides, SC-LMKB on ICFG-PEDES dataset can achieve 28.0\%, 64.8\%, and 19.5\% mAP gains compared to the same baselines when SNR is 25 dB as shown in Fig.~\ref{figUMi}\subref{figUMi:sub2}. Fig.~\ref{figUMi}\subref{figUMi:sub3} shows that SC-LMKB on the RSTPReid dataset can achieve 36.9\%, 79.6\%, and 18.1\% mAP gains compared to the same baselines when SNR is 25 dB. The mAP improvement demonstrates that the SC-LMKB can utilize LMKB to improve system performance over the MIMO channel.

\begin{figure}[t]
    \centering
    \subfloat[Rank@1]{%
        \includegraphics[width=0.48\linewidth]{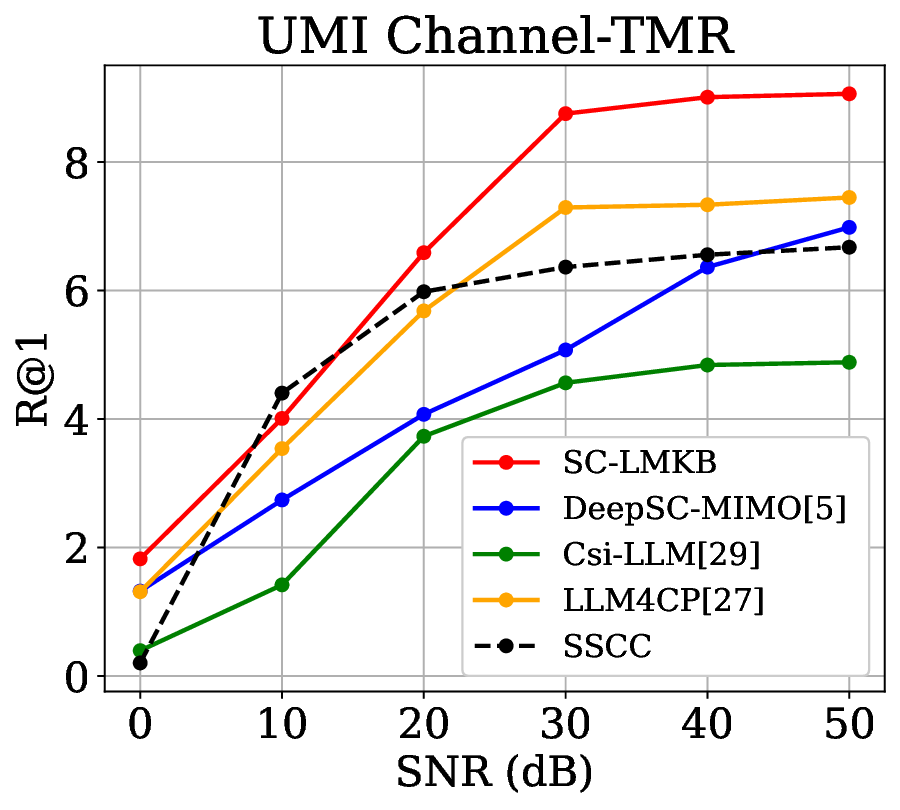}
        \label{TMR:sub1}
    }
    \subfloat[Rank@5]{%
        \includegraphics[width=0.48\linewidth]{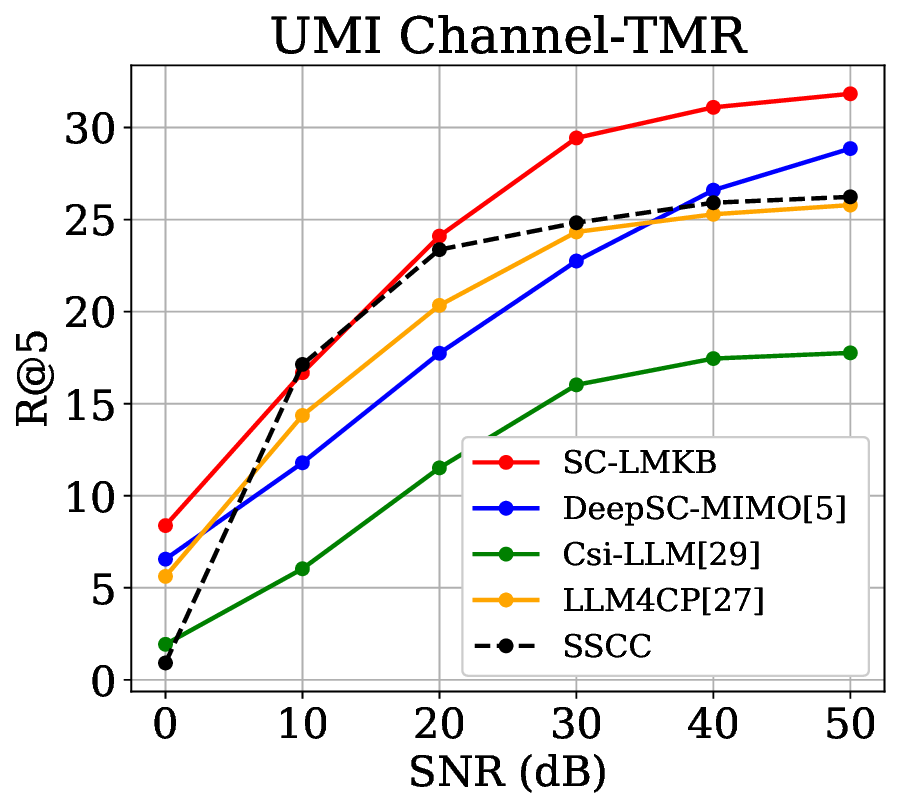}
        \label{TMR:sub2}
    }\\[0.5ex]
    \subfloat[Rank@10]{%
        \includegraphics[width=0.48\linewidth]{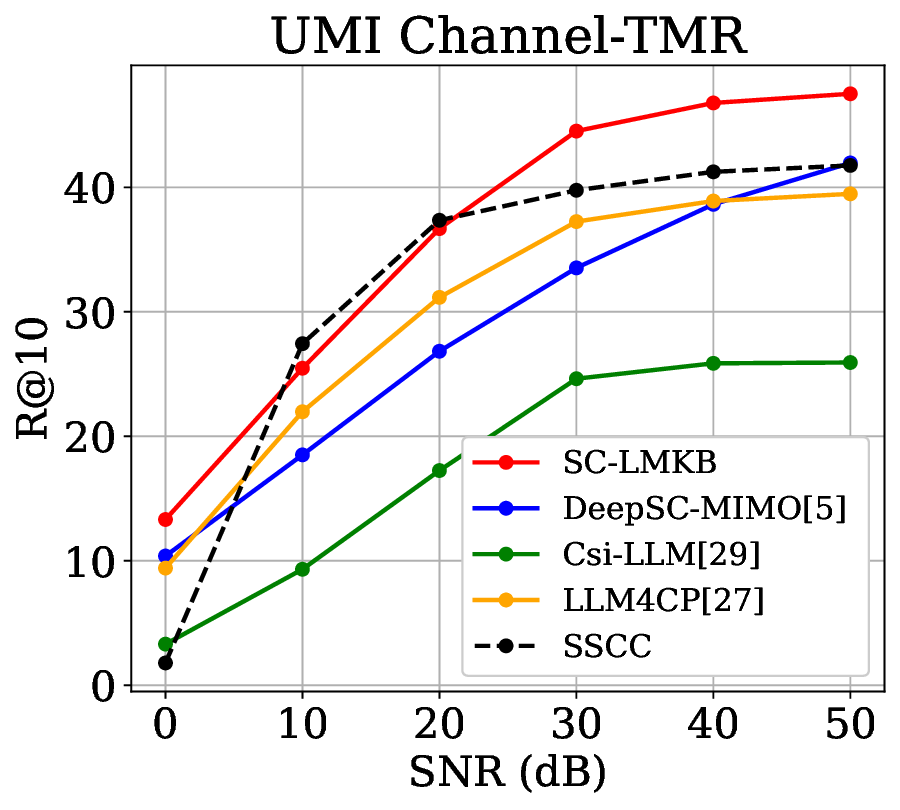}
        \label{TMR:sub3}
    }
    \caption{Performance under different SNR levels over the UMi-LOS channel on the TMR dataset.}
    \label{TMR}
\end{figure}
\begin{figure}[t]
    \centering
    \subfloat[Rank@1]{%
        \includegraphics[width=0.48\linewidth]{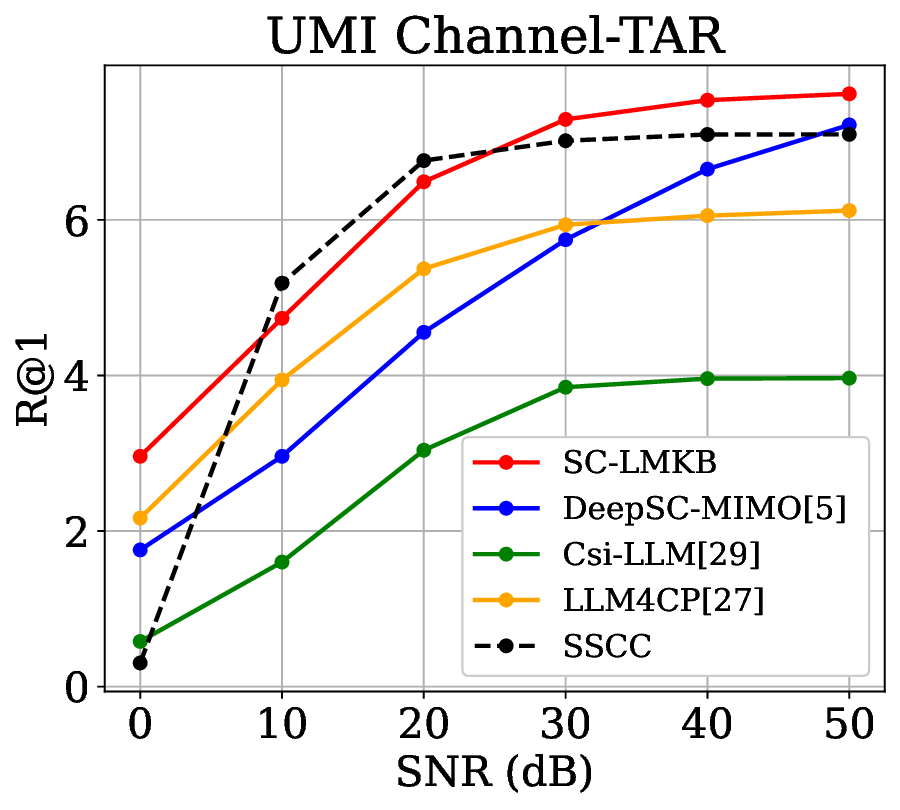}
        \label{TAR:sub1}
    }
    \subfloat[Rank@5]{%
        \includegraphics[width=0.48\linewidth]{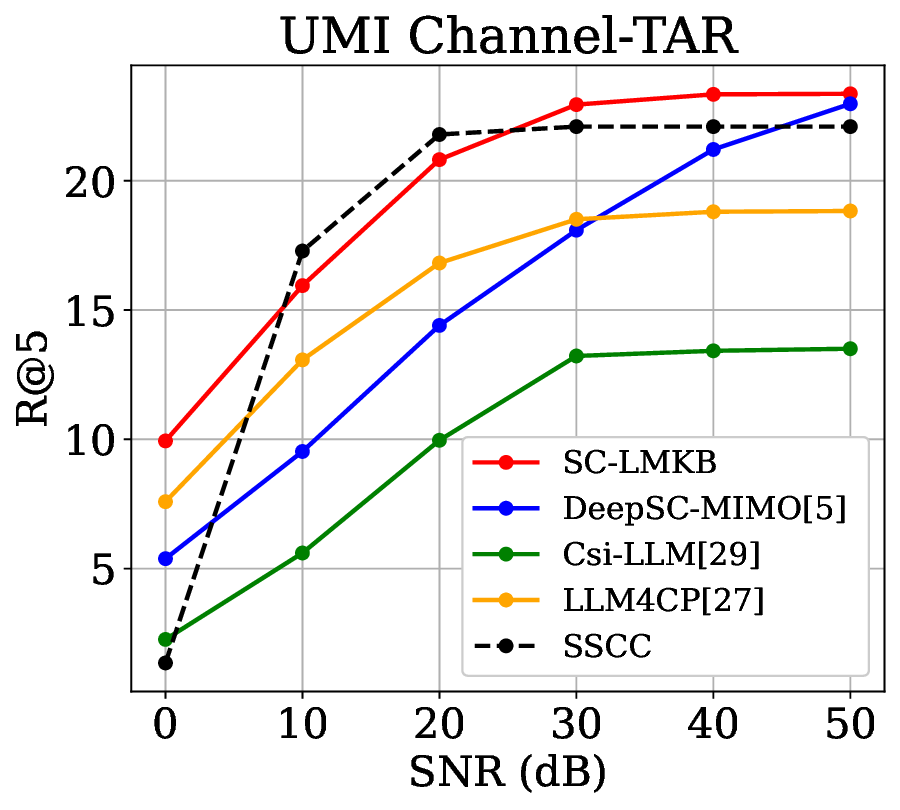}
        \label{TAR:sub2}
    }\\[0.5ex]
    \subfloat[Rank@10]{%
        \includegraphics[width=0.48\linewidth]{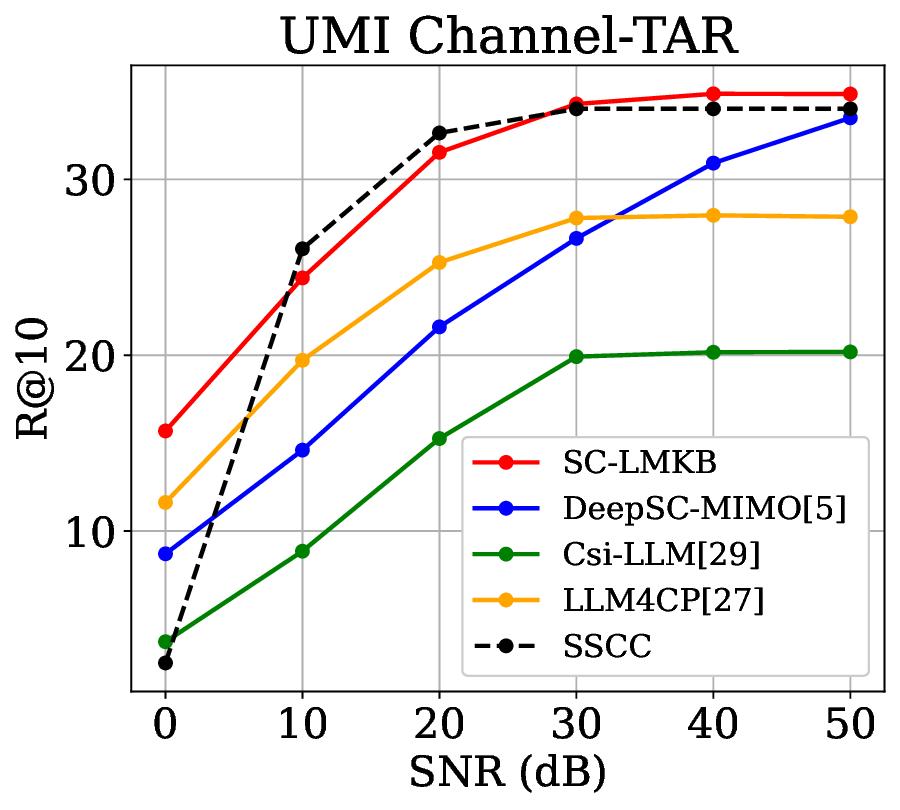}
        \label{TAR:sub3}
    }
    \caption{Performance under different SNR levels over the UMi-LOS channel on the TAR dataset.}
    \label{TAR}
\end{figure}

\textbf{Improvements on the TMR dataset.} Fig. \ref{TMR} shows the performance gains of all systems with SNRs over the UMi-LOS channel on the adopted TMR dataset. As shown in Fig.~\ref{TMR}\subref{TMR:sub1}, it can be observed that the Rank@1 increases with SNR and gradually converges to a certain threshold. This is because higher SNR means a better communication condition and can decrease transmission errors, which consequently increases the task performance at the receiver. It can be observed that the proposed SC-LMKB outperforms the LLM4CP, Csi-LLM, DeepSC-MIMO and the traditional method across the entire SNR region. This demonstrates the effectiveness of our proposed LMKB.  Specifically, when SNR is 30 dB, SC-LMKB achieves improvements of 72.6\%, 90.7\%, and 20.0\% Rank@1 gains compared to DeepSC-MIMO, Csi-LLM, and LLM4CP, respectively.
The results in Fig.~\ref{TMR}\subref{TMR:sub2} and Fig.~\ref{TMR}\subref{TMR:sub3} exhibit the similar trends in Fig.~\ref{TMR}\subref{TMR:sub1}. 
However, the traditional method has a slightly better performance at an SNR of 10 dB as shown in Fig. \ref{TMR} due to the error correction capability of conventional channel coding.
Besides, SC-LMKB can achieve 29.5\%, 83.7\%, and 21.0\% Rank@5 gains compared to the same baselines when SNR is 30 dB as shown in Fig.~\ref{TMR}\subref{TMR:sub2}. Fig.~\ref{TMR}\subref{TMR:sub3} shows that SC-LMKB can achieve 32.8\%, 80.9\%, and 19.6\% Rank@10 gains compared to the same baselines when SNR is 30 dB. The performance improvement demonstrates that the SC-LMKB can utilize LMKB to improve system performance over the MIMO channel.

\textbf{Improvements on the TAR dataset.} Fig. \ref{TAR} shows the performance gains of all systems with SNRs over the UMi-LOS channel on the adopted TAR dataset. As shown in Fig.~\ref{TAR}\subref{TAR:sub1}, it can be observed that the Rank@1 increases with SNR and gradually converges to a certain threshold. This is because higher SNR means a better communication condition and can decrease transmission errors, which consequently increases the task performance at the receiver. It can be observed that the proposed SC-LMKB outperforms the LLM4CP, Csi-LLM, DeepSC-MIMO and the traditional method across the entire SNR region. This demonstrates the effectiveness of our proposed LMKB. Specifically, when SNR is 30 dB, SC-LMKB achieves improvements of 26.9\%, 89.5\%, and 22.2\% Rank@1 gains compared to DeepSC-MIMO, Csi-LLM, and LLM4CP, respectively.
The results in Fig.~\ref{TAR}\subref{TAR:sub2} and Fig.~\ref{TAR}\subref{TAR:sub3} exhibit the similar trends in Fig.~\ref{TAR}\subref{TAR:sub1}. 
However, the traditional method has a slightly better performance at an SNR of 10 dB as shown in Fig. \ref{TAR} because low SNRs will affect the performance of the effect of CDG and CDFC, thus leading to performance degradation.
Besides, SC-LMKB can achieve 26.8\%, 73.5\%, and 24.0\% Rank@5 gains compared to the same baselines when SNR is 30 dB as shown in Fig.~\ref{TAR}\subref{TAR:sub2}. Fig.~\ref{TAR}\subref{TAR:sub3} shows that SC-LMKB can achieve 28.7\%, 72.1\%, and 23.3\% Rank@10 gains compared to the same baselines when SNR is 30 dB. The performance improvement demonstrates that the SC-LMKB can utilize LMKB to improve system performance over the MIMO channel.
\begin{figure*}[htbp]
    \centering
    \subfloat[CUHK-PEDES]{
        \includegraphics[width=0.31\textwidth]{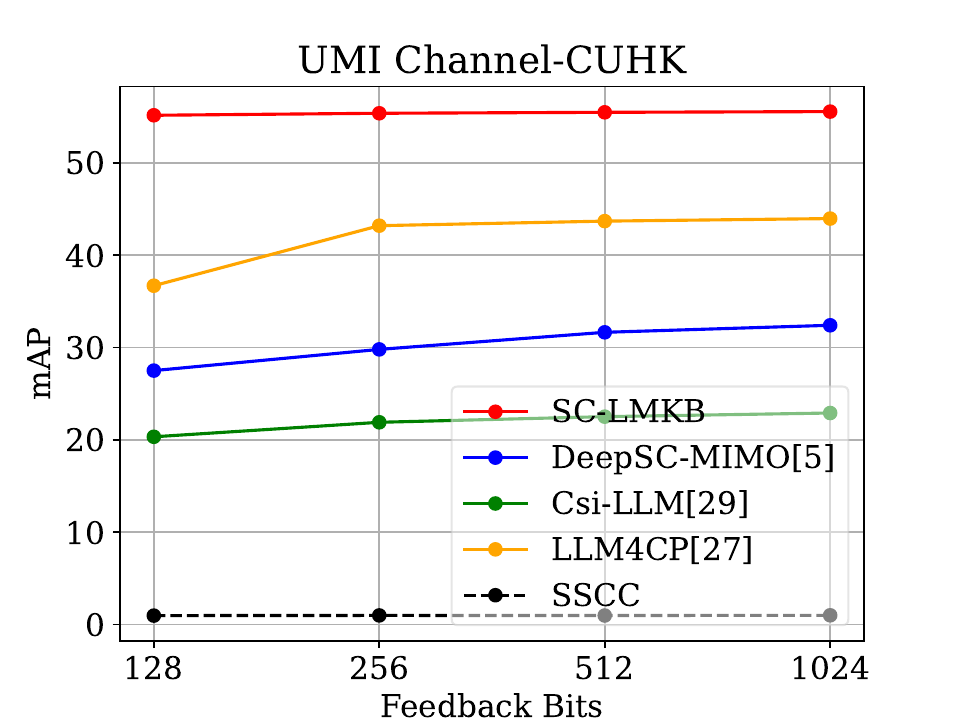}
        \label{figBITS:sub1}
    }
    \hfill
    \subfloat[ICFG-PEDES]{
        \includegraphics[width=0.31\textwidth]{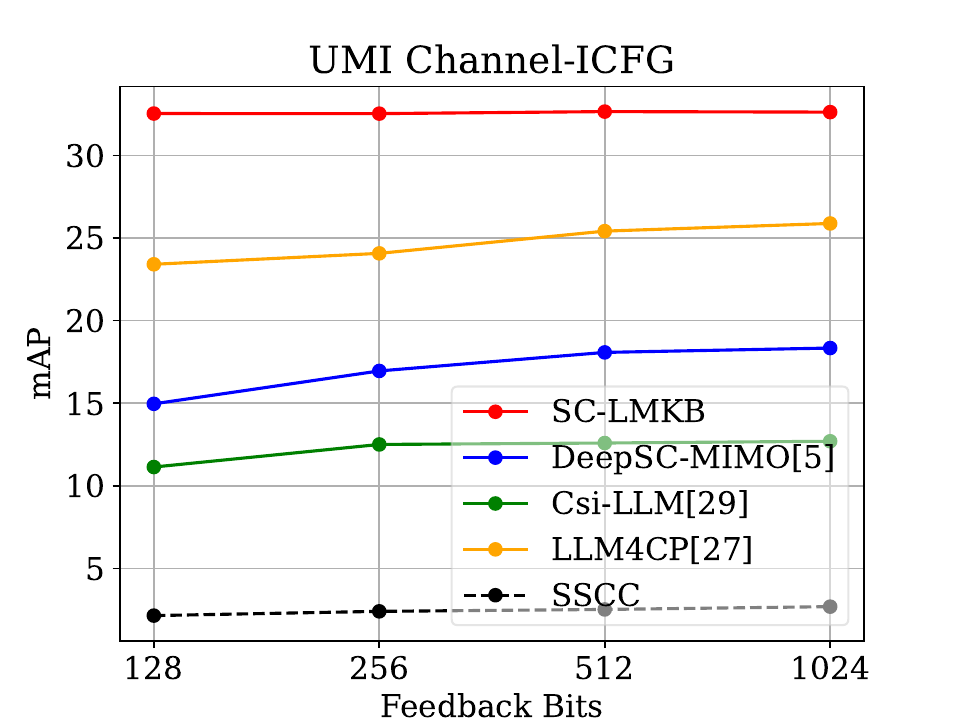}
        \label{figBITS:sub2}
    }
    \hfill
    \subfloat[RSTPReid]{
        \includegraphics[width=0.31\textwidth]{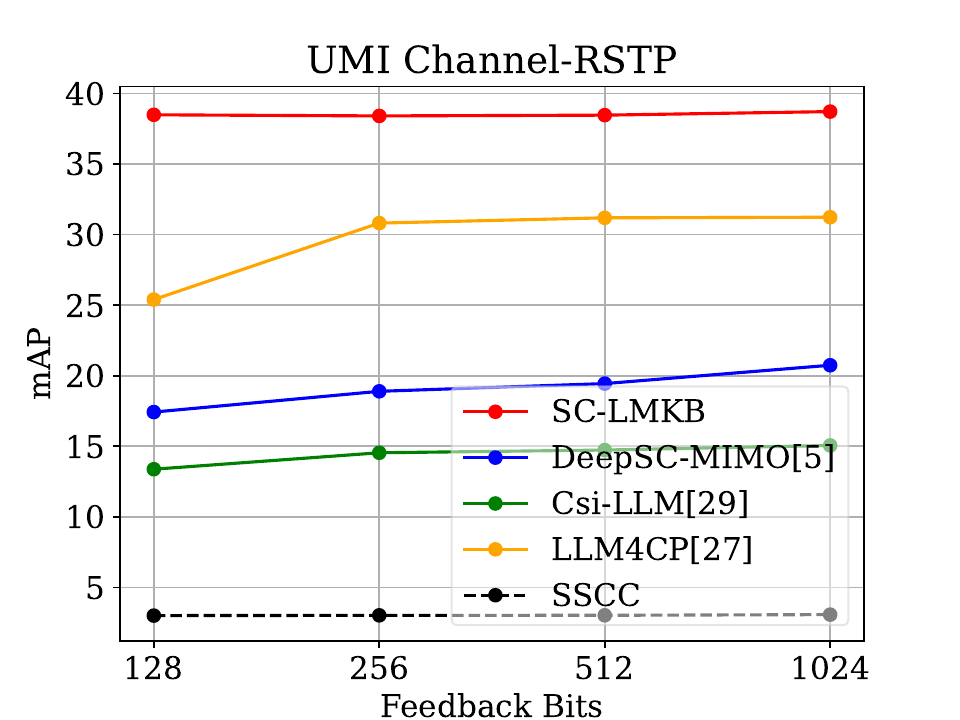}
        \label{figBITS:sub3}
    }
    \caption{mAP versus feedback bits over UMi-LOS channel on TPR datasets.}
    \label{figBITS}
\end{figure*}
\begin{figure*}[htbp]
    \centering
    \subfloat[CUHK-PEDES]{
        \includegraphics[width=0.28\textwidth]{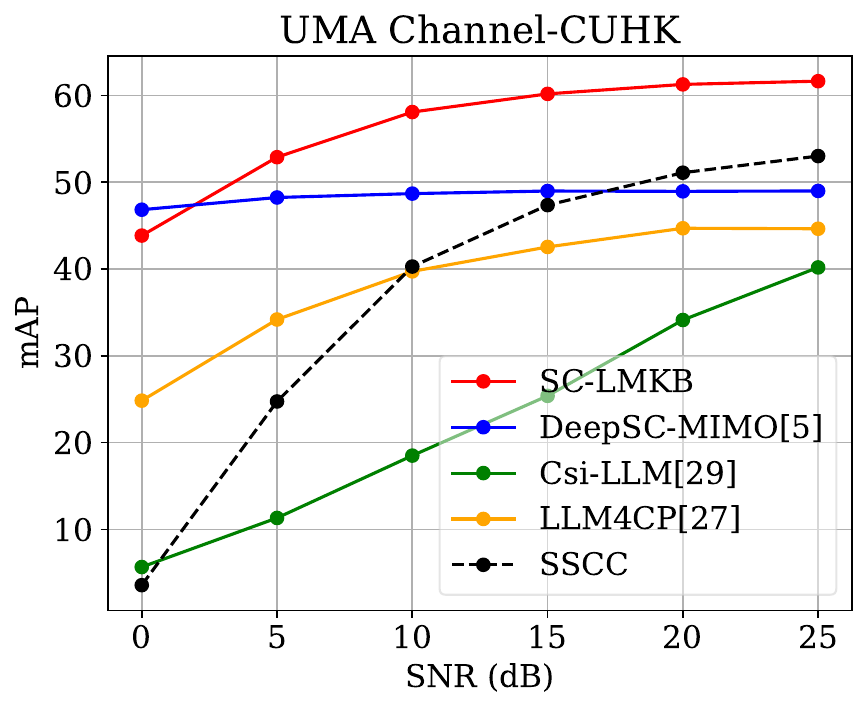}
        \label{figUMa:sub1}
    }
    \hfill
    \subfloat[ICFG-PEDES]{
        \includegraphics[width=0.28\textwidth]{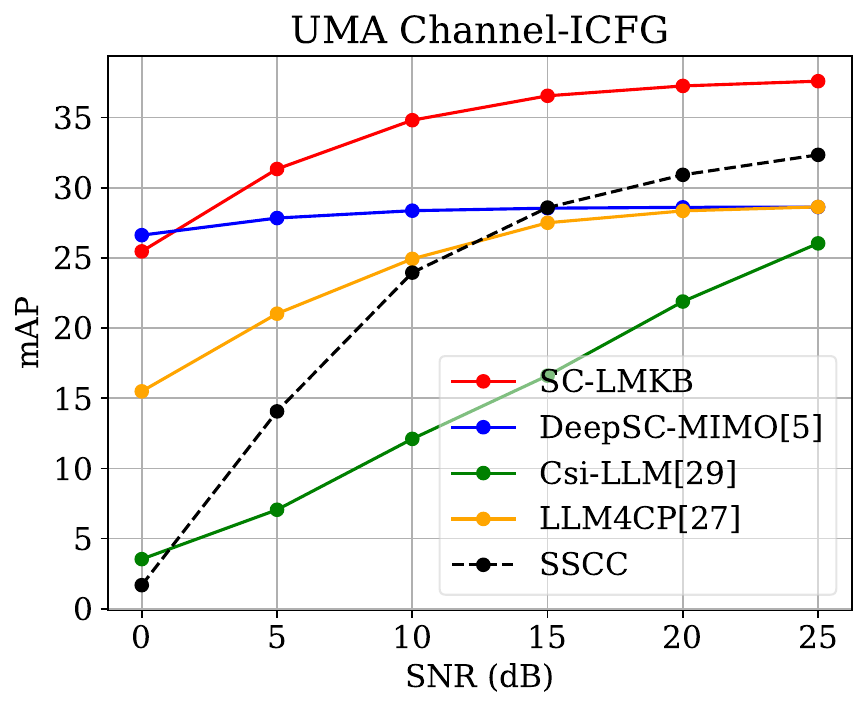}
        \label{figUMa:sub2}
    }
    \hfill
    \subfloat[RSTPReid]{
        \includegraphics[width=0.28\textwidth]{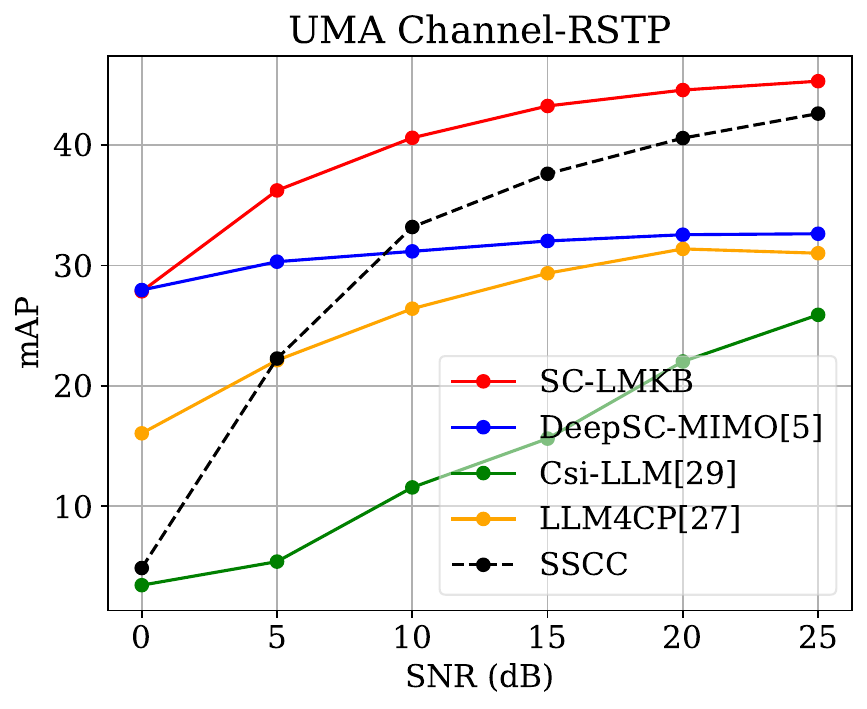}
        \label{figUMa:sub3}
    }
    \caption{mAP versus SNR over UMa-NLOS channel on TPR datasets.}
    \label{figUMa}
\end{figure*}
Furthermore, we compare the proposed SC-LMKB with baseline models under varying numbers of CSI feedback bits over the UMi-LOS channel. Fig.~\ref{figBITS} shows the mAP of all evaluated systems across different feedback bit levels on the three adopted TPR datasets.
% As shown in Fig.~\ref{figBITS}\subref{figBITS:sub1}, it can be observed that the mAP increases with feedback bits and gradually converges to a certain threshold. This is because higher feedback resolution leads to more accurate CSI reconstruction, which reduces transmission errors and ultimately enhances task performance at the receiver.
It can be observed that the proposed SC-LMKB outperforms all baselines across the entire feedback bits region. This performance advantage stems from the fact that methods like DeepSC-MIMO and traditional approaches rely heavily on accurate CSI feedback, while SC-LMKB only requires partial CSI information and is capable of predicting future CSI based on noisy historical data. Therefore, SC-LMKB exhibits lower sensitivity to feedback bandwidth and maintains superior performance even with limited feedback.
In particular, when the number of feedback bits is 128, SC-LMKB achieves a performance gain of 97.1\% in mAP compared to DeepSC-MIMO. Similar trends can be observed in Fig.~\ref{figBITS}\subref{figBITS:sub2} and  Fig.~\ref{figBITS}\subref{figBITS:sub3}.
Besides, SC-LMKB on ICFG-PEDES dataset can achieve 101.8\% mAP gains compared to the DeepSC-MIMO when the feedback bit is 128 bit, as shown in Fig.~\ref{figBITS}\subref{figBITS:sub2}. Fig.~\ref{figBITS}\subref{figBITS:sub3} shows that SC-LMKB on the RSTPReid dataset can achieve 89.0\% mAP gains compared to DeepSC-MIMO when the feedback bit is 128 bits. The mAP improvement demonstrates that the SC-LMKB is robust over feedback bits.

To evaluate the generalization ability of the proposed LMKB, we directly apply the model trained in the UMi-LOS channel to the UMa-NLOS channel without additional training process. Fig. \ref{figUMa} shows the mAP of all systems with SNRs over the UMa-NLOS channel on the adopted three TPR datasets. The results exhibit similar trends to those under the UMi-LOS channel, illustrating the generalization of the proposed LMKB. Specifically, when SNR is 25 dB, Fig.~\ref{figUMa}\subref{figUMa:sub1} shows SC-LMKB can achieve up to 25.5\%, 53.3\%, and 38.1\% mAP gains compared to DeepSC-MIMO, Csi-LLM, and LLM4CP, respectively.
Fig.~\ref{figUMa}\subref{figUMa:sub2} shows that SC-LMKB on ICFG-PEDES dataset can achieve 31.4\%, 44.4\%, and 31.3\% mAP gains compared to the same baselines when SNR of the UMa-NLOS channel is 25 dB. 
Fig.~\ref{figUMa}\subref{figUMa:sub3} shows that SC-LMKB on RSTPReid dataset can achieve 38.9\%, 74.9\%, and 46.0\% mAP gains compared to the same baselines at the same channel condition as in Fig.~\ref{figUMa}\subref{figUMa:sub2}. 

\begin{figure}[htbp]
\centering
    \subfloat[UMi-LOS]{%
        \includegraphics[width=0.9\linewidth]{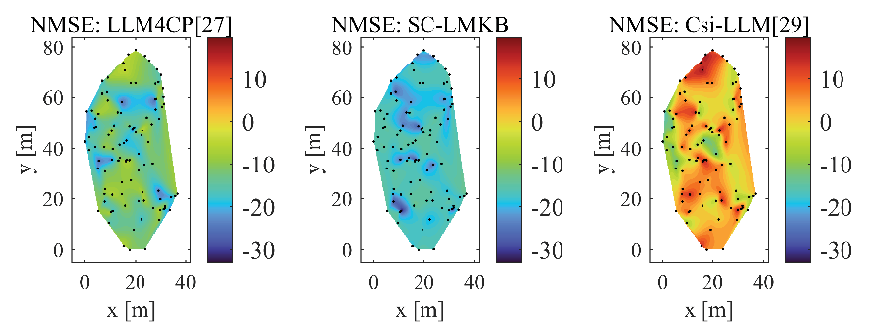}
        \label{fig:umi}
    }\\[1ex] % 换行并添加间距
    \subfloat[UMa-NLOS]{%
        \includegraphics[width=0.9\linewidth]{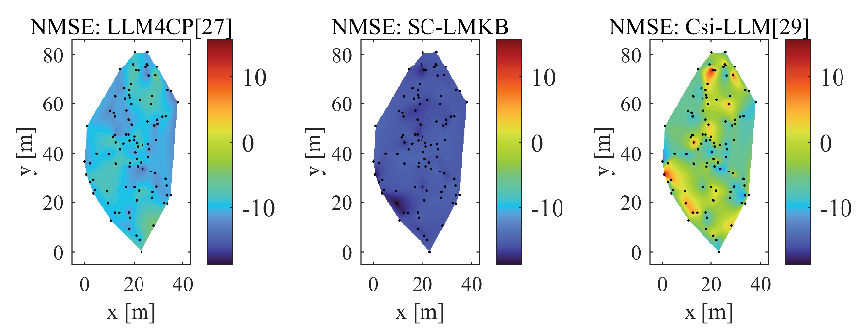}
        \label{fig:uma}
    }
    \caption{Visualization results of the proposed SC-LMKB over (a) UMi-LOS and (b) UMa-NLOS channels.}
    \label{fig:channel_visualization}
\end{figure}

% \begin{figure*}[t!]
%     \centering
%     \subfloat[UMi-LOS]{%
%         \includegraphics[width=0.8\linewidth]{UMi-LOS.eps}
%         \label{fig:umi}
%     }\\[1ex] % 换行并添加间距
%     \subfloat[UMa-NLOS]{%
%         \includegraphics[width=0.8\linewidth]{UMa-NLOS.eps}
%         \label{fig:uma}
%     }
%     \caption{Visualization results of the proposed SC-LMKB over (a) UMi-LOS and (b) UMa-NLOS channels.}
%     \label{fig:channel_visualization}
% \end{figure*}

Fig.~\ref{fig:channel_visualization} shows the visualization results of the proposed SC-LMKB method, along with existing LLM-enabled CSI generation baselines, namely Csi-LLM and LLM4CP. The visualization covers a fan-shaped area containing 100 user terminals. The color intensity in the figure reflects the NMSE between the predicted and ground-truth CSI values where darker colors indicate higher NMSE, while lighter colors denote better prediction accuracy. As shown in FIg. \ref{fig:channel_visualization}, across both 3GPP UMi-LOS and UMa-NLOS scenarios, the proposed SC-LMKB consistently achieves lower NMSE compared to other methods. This observation further validates the higher mAP gains. In particular, the improvements under the UMa-NLOS setting demonstrate the generalization capability of SC-LMKB over unseen and more complex propagation environments, outperforming existing LLM-enabled approaches.
\begin{figure}[htbp]
\centering
\includegraphics[width=0.7\linewidth]{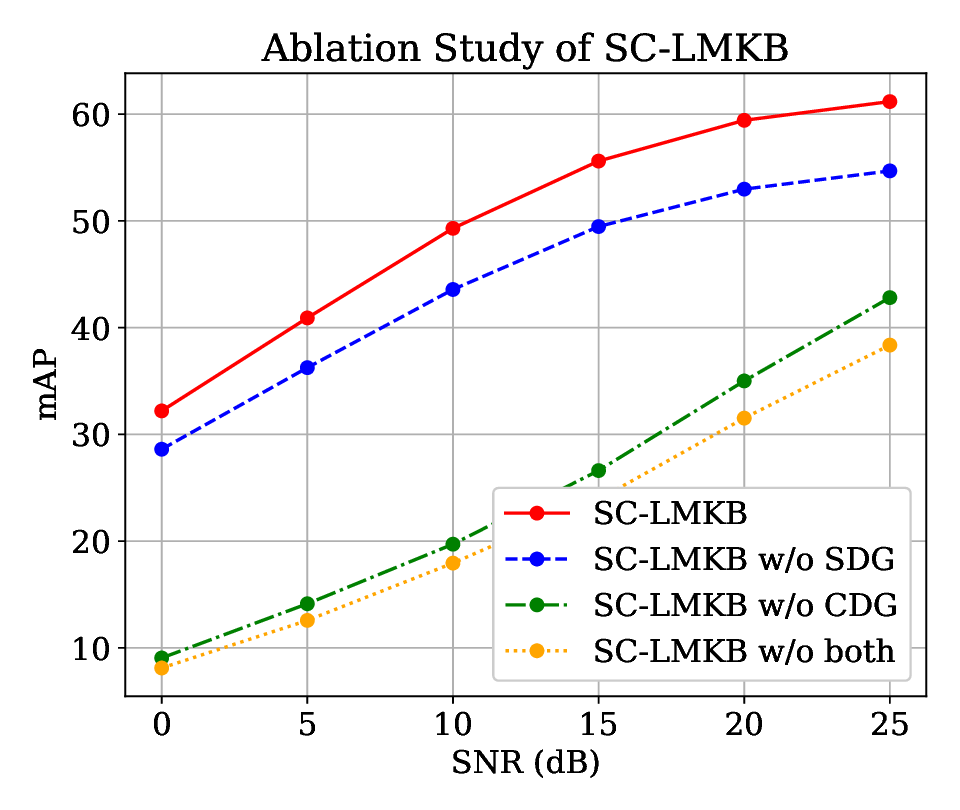}
\caption{Ablation experiments.}
\label{ablation}
\end{figure}

\subsection{Ablation Experiments}

To further evaluate the contribution of each component in SC-LMKB, we conduct ablation experiments by selectively disabling different data generation modules. Specifically, we consider the following variants:
\begin{itemize}
    \item \textbf{SC-LMKB w/o SDG}: The SDG module with CDFC framework is removed.
    \item \textbf{SC-LMKB w/o CDG}: The CDG module is removed.
    \item \textbf{SC-LMKB w/o both}: Both the SDG and CDG modules are removed.
\end{itemize}

Fig.~\ref{ablation} illustrates the comparative performance of the ablated SC-LMKB variants under different SNR values over the UMi-LOS channel. The results demonstrate that SDG and CDG contribute significantly to the overall performance of SC-LMKB. The complete model consistently outperforms all ablated versions across various SNR levels. In particular, at an SNR of 15~dB, SC-LMKB achieves an improvement of 12.5\% over SC-LMKB w/o SDG and 109\% over SC-LMKB w/o CDG. This indicates that the CDG component has a more pronounced impact on performance compared to SDG, highlighting the critical role of accurate and diverse channel data in MIMO SC.

\section{Conclusion}
In this paper, we have proposed a SC-LMKB, which has integrated an SC system with an LMKB. 
%To address the data limitation issue, an LLM-enabled generation mechanism has been proposed to leverage the LMKB to generate data.
In particular, a prompt engineering strategy has been proposed to generate source data. Besides, for the channel data, a cross-attention alignment method has been proposed to align CSI features with the natural language modality in the LLM space. Then, to resist the semantic
noise induced by hallucination from LLMs, a CDFC framework has been proposed to alleviate the hallucination in SDG. 
%In particular, the framework first filters out the data that deviates from the intended source based on semantic similarity and then fuses source data with filtered LLM-generated data based on semantic importance. 
Besides, a joint training objective that combines
cross-entropy loss and reconstruction loss has been proposed to reduce the impact of hallucination on CDG. Experimental results have demonstrated that the proposed SC-LMKB system effectively utilizes the LLM to generate additional data and then enhances the task performance. Our work provides a novel perspective on SC systems enabled with LLMs and offers a promising solution for the hallucination problem. Future research will explore the application to different source modalities and adaptation to other fundamental models.

\bibliographystyle{IEEEtran}
\bibliography{citationnew}

\vfill

\end{document}